\documentclass[traditabstract]{aa}
\usepackage{txfonts}
\usepackage{graphicx}
\usepackage{subfigure}
\usepackage{natbib}
\bibpunct{(}{)}{;}{a}{}{,} 

\begin{document}
\title{A simple prescription for simulating and characterizing gravitational arcs}
\author{Cristina Furlanetto\inst{1,2}
\and Bas\'ilio X. Santiago\inst{1,2}
\and Mart\'in Makler\inst{3,2}
\and Cl\'ecio de Bom\inst{3,2}
\and Carlos H. Brandt\inst{4,3}
\and Angelo Fausti Neto\inst{2}
\and Pedro C. Ferreira\inst{5}
\and Luiz Nicolaci da Costa\inst{6,2}
\and Marcio A. G. Maia\inst{6,2}
 }
\institute{Departamento de Astronomia, Universidade Federal do Rio Grande do Sul, Av. Bento Gon\c{c}alves, 9500, Porto Alegre, RS 91501-970, Brazil
\and Laborat\'orio Interinstitucional de e-Astronomia, Rua Gen. Jos\'e Cristino, 77, Rio de Janeiro, RJ 20921-400, Brazil
\and Centro Brasileiro de Pesquisas F\'isicas, Rua Dr. Xavier Sigaud 150, Rio de Janeiro, RJ 22290-180, Brazil
\and Laborat\'orio Nacional de Computa\c c\~ao Cient\'ifica, Av. Get\'ulio Vargas, 333, Petr\'opolis, RJ 25651-075, Brazil
\and Departamento de F\'isica, 
Universidade Federal do Rio Grande do Norte, Campus Universit\'ario, 
Natal, RN  59072-970, Brazil
\and Observat\'orio Nacional, Rua Gen. Jos\'e Cristino, 77, Rio de Janeiro, RJ 20921-400, Brazil
}
\abstract{Simple models of gravitational arcs are crucial 
for simulating large samples of these objects with full control of
the input parameters. These models also provide approximate and automated 
estimates of the shape and structure of the arcs, which 
are necessary for detecting and characterizing these objects on
massive wide-area imaging surveys. 
We here present and explore the ArcEllipse, a simple prescription for 
creating objects with a shape similar to gravitational arcs. 
We also present PaintArcs, which is a code that couples this geometrical 
form with a brightness distribution and adds the resulting object to images. 
Finally, we introduce
ArcFitting, which is a tool that fits ArcEllipses to images of real 
gravitational arcs. We validate this fitting technique using simulated arcs
and apply it to CFHTLS and HST images of tangential arcs around clusters
of galaxies. Our simple ArcEllipse model for the arc, associated to a S\'ersic 
profile for the source, recovers the total signal in real images typically within 10\%-30\%. The ArcEllipse+S\'ersic models also automatically 
recover visual estimates of length-to-width ratios of real arcs. Residual
maps between data and model images reveal the incidence of arc substructure. 
They may thus be used as a diagnostic for arcs formed by the
merging of multiple images. The incidence of these substructures is the
main factor that prevents ArcEllipse models from accurately describing
real lensed systems.}

\maketitle

\section{Introduction}

Gravitational lensing is an important tool for a variety of astrophysical 
applications. In particular, gravitationally lensed arcs 
produced by massive galaxy clusters can provide information on their total 
inner mass distribution \citep{1999PThPS.133....1H,2002MNRAS.335..311G,2006AAS...208.6805M,2010ARA&A..48...87T,2011A&ARv..19...47K}. Statistical analyses of gravitational arcs provide
constraints on the evolution of structure with cosmic time and on the
cosmological parameters \citep{2007MNRAS.382..121H,2007A&A...464..845V,2006ApJ...640...47K,2004MPLA...19.1083M,2002A&A...387..788G,1998A&A...330....1B,2007ApJ...660.1176E,2008AJ....135..664H,Kausch2010,2012ApJ...749...38M}. 
Because of the flux amplification associated to 
gravitational lensing, these arcs are also a probe for the high-redshift 
universe, which is unaccessible through unlensed sources. 

On the other hand, gravitational arcs are very rare objects, since each 
square degree of the sky contains approximately one galaxy cluster massive enough 
to produce arcs by gravitational lensing \citep{2005ApJS..157....1G}. So far an order of 
hundreds of these objects were detected \citep{Luppino1999,Zaritsky2003,2003ApJ...593...48G,2005ApJ...627...32S,2007A&A...461..813C,2008ApJ...682..964B}. 
Future wide-field surveys with sub-arcsecond seeing, such as the Dark Energy 
Survey (DES)\footnote{http://www.darkenergysurvey.org/}, will provide large 
samples of gravitational arc systems that will enable numerous statistical studies. 
Because of the large area, it will be virtually impossible to visually inspect the 
whole survey to look for gravitationally lensed sources. Therefore, automated 
detection methods are ultimately needed.

There are a few proposals for automated detection algorithms in the literature, 
although none of them is considered as the default algorithm \citep[e.g.,][]{2004A&A...416..391L,2005ApJ...633..768H,2007A&A...472..341S,2007A&A...461..813C,2012ApJ...749...38M,mediatrix}. 
Moreover, most of these algorithms are not fully characterized in the sense that 
they do not have a determined efficiency and contamination. These 
quantities could be quantified by using samples of images with simulated 
arcs, which would act as truth tables for this type of study. 
Furthermore, simulated arc images are useful to test in controlled studies 
arc measurement tools and image pre-processing algorithms used to enhance 
their detectability.

Currently, the literature lacks sufficiently detailed works related to the fast simulation
and characterization of gravitational arcs. 
Arc simulations have mostly
been based on ray-tracing techniques given a distribution of source
and lens properties. The common practice is to
measure the arc's length $L$, width $W$, and its distance 
to the center of the lens. 
Few authors have tried to constrain their shapes and orientation. 
\citet{Miralda1993} introduced the idea of measuring $L$ by fitting a circle
arc through the arc center and its extremes. 
\citet{Bartelmann1994} have pioneered the idea of approximating gravitational 
arcs as an ellipse whose major axis is equal to $L$ (the so-called {\it ellipse
fitting}). However, we are not aware of any preceding study that has
introduced a surface brightness distribution to simulate and characterize
gravitational arcs. 

Motivated by these questions, we developed a code called PaintArcs which 
draws geometric figures that mimic arcs and adds them to images. It currently 
uses the prescription of an ArcEllipse, which is an analytical expression 
for the 
deformation of an ellipse such that one of its main axis becomes a circle 
segment. Only a handful of fully controllable parameters go into this
simple prescription. We use the ArcEllipse in conjunction 
with a S\'ersic profile to take into 
account the surface brightness distribution of the simulated arc. 
We also present ArcFitting, which is a code that uses the ArcEllipse+S\'ersic model to estimate the morphological and structural parameters of real arcs.

The prescriptions that we adopt here for the shape and structure
of arcs are certainly a simplification of real gravitational arcs, which
are often formed by the merging of individual multiple images of a source. One
such situation are the giant arcs formed when a source is close to the
cusp of a caustic. Particularly when real arcs are imaged in high resolution,
the ArcEllipse shape with a S\'ersic profile will not accommodate
their substructures. Our goal here is to exactly to quantify the 
observed discrepancies from a simple model and, as a result, assess which 
improvements are most promising and to what extent
these simple models can be used in controlled experiments for automatic
arc detection algorithms.

This paper is organized as follows. In \S \ref{arcellipse} we present
the ArcEllipse, including the addition of a surface brightness
profile and asymmetry. In \S \ref{painting} we describe the PaintArcs code, accounting for image pixelation, and including 
seeing effects and Poisson noise in the simulated image. 
The ArcFitting methods are presented in \S \ref{arcfitting} and their
application to fit real images are presented in \S 
\ref{fitapplication}. Finally, in \S \ref{conclude}, we discuss the results and future work.

\section{ArcEllipse}\label{arcellipse}

The ArcEllipse consists of a simple idea to analytically express arc shapes as the distortion of an ellipse such that one of its main axes is bent into an arc of a circle. The resulting geometrical figure can be used to create a surface brightness map that mimics a gravitational arc. 

An ordinary ellipse is formed by the set of points whose distances to the x and y axes, weighted by the semi-axes, $a$ and $b$, satisfy 
\begin{equation}
\left(\frac{x}{a}\right)^2 +\left(\frac{y}{b}\right)^2 = 1.
\label{elipse-equation}
\end{equation}

Following the same idea, we can define the ArcEllipse as the set of points whose distances from a point on the circumference along the tangential direction ($r_c \,\Delta \theta$) and along
the radial direction ($\Delta r$) satisfy
\begin{equation}
\left(\frac{ r_c \,\Delta \theta }{a}\right)^2 + \left(\frac{\Delta r }{b}\right)^2 = 1,
\label{arcellipse-equation}
\end{equation}
where $r_c$ is the radius of curvature of the circle (and thus the curvature radius of the arc).

\begin{figure*}[!htb]
\begin{center}
\begin{minipage}[b]{0.9\linewidth}
\includegraphics[scale=0.4]{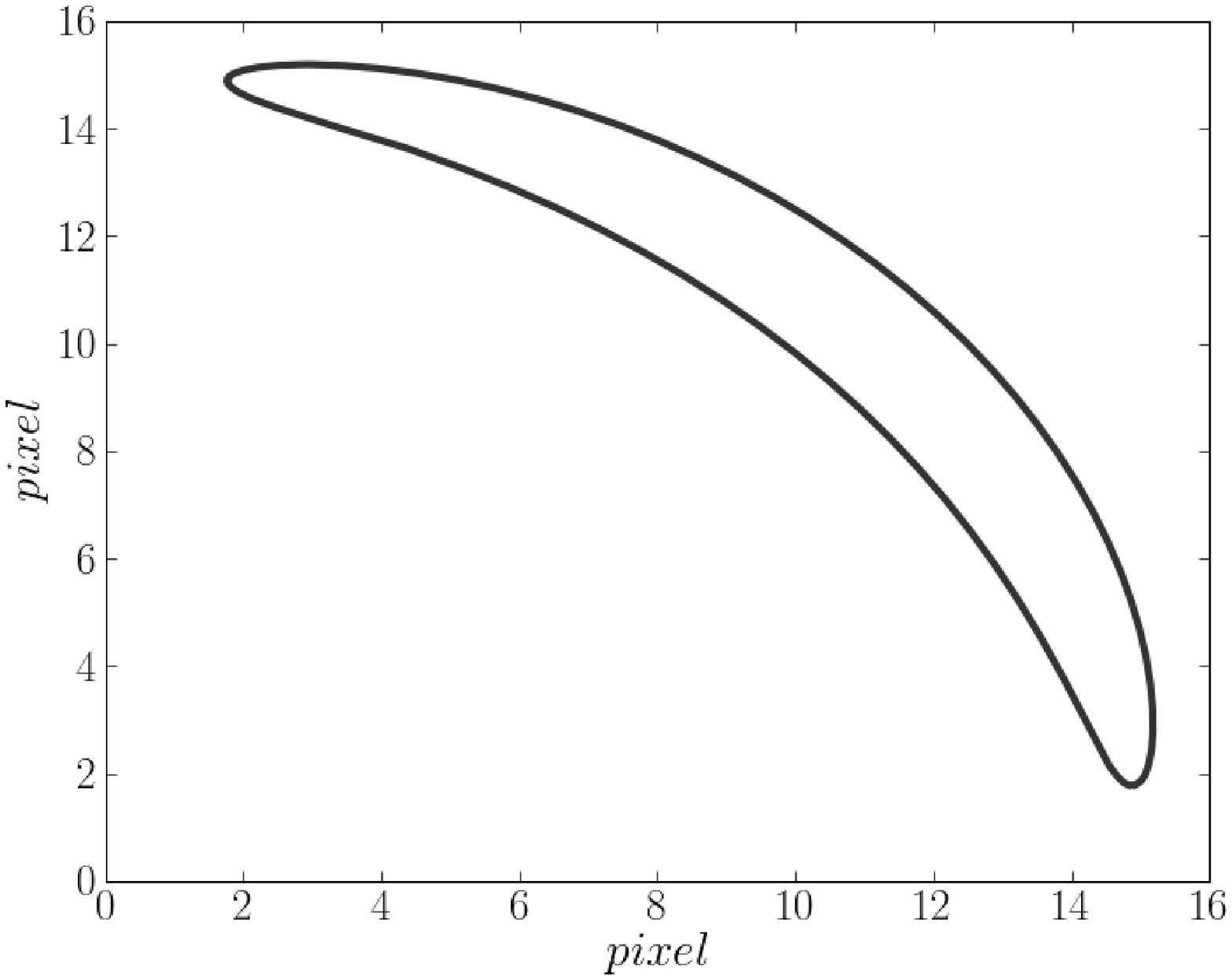}
\includegraphics[scale=0.4]{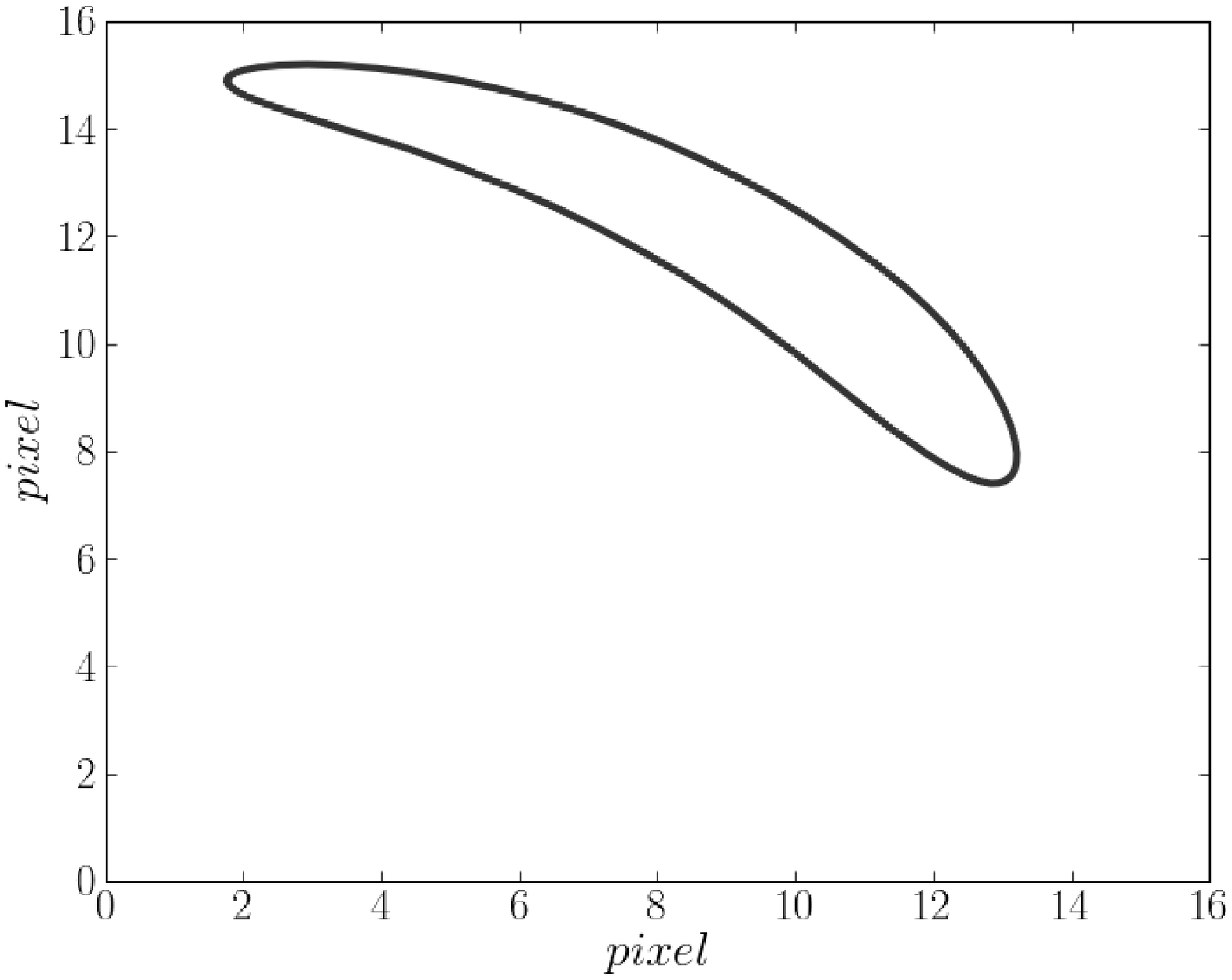}
\caption{Geometrical figures produced by the ArcEllipse. Left panel: symmetric ArcEllipse created with $e = 1 - b/a = 0.9$,  $b = 1$ pixel, $r_c = 15$ pixels, and $\theta_0 = 45^o$. 
Right panel: asymmetric ArcEllipse created with $e_1 = 0.9, e_2 = 0.75$, 
with the remaining parameters as in the left panel.}
\label{fig_arcellipse}
\end{minipage}
\end{center}
\end{figure*}

This simple expression indeed gives a shape that resembles an arc, as we can see in the left panel of Fig. \ref{fig_arcellipse}.  
We define the length of an ArcEllipse as the length of the arc segment between the two extreme points where $\Delta r = 0$. This leads to $L = 2 a$. The width at $\Delta \theta =0 $ is given by $W=2 b$.
Therefore, the length-to-width ratio of the ArcEllipse 
is simply given as $L/W=a/b$.

In polar coordinates, if the origin of the coordinate system corresponds to the center of curvature of the ArcEllipse, we have $\Delta r = r-r_c$ and $\Delta \theta = \theta-\theta_0$, where $\theta_0$ is the position angle 
of the ArcEllipse center, such that
\begin{equation}
r_{\pm} = r_c \pm b \sqrt{1-\left[\frac{r_c \left(\theta_0 - \theta \right)}{a} \right]^2}.
\label{r_arcellipse}
\end{equation}
\mbox{}

The ArcEllipse is therefore defined by four parameters: the length $2a$, the width $2b$, the curvature radius $r_c$, and the position angle $\theta_0$.

From the solution above, it is simple to show that the area of the ArcEllipse is given by
\begin{eqnarray}
A_{\rm ArcEll} &=&\int^{\theta_0 + a/r_c}_{\theta_0 - a/r_c} \int^{r_+(\theta)}_{r_-(\theta)} r \, dr \, d\theta  = \nonumber \\ 
&=& \int^{\theta_0 + a/r_c}_{\theta_0 - a/r_c} \frac{\left(r_+(\theta)^2\right)-\left(r_-(\theta)^2\right)}{2}  d\theta = \nonumber \\
&=& 
 \pi \, a \, b = \frac{\pi \,L \,W}{ 4},
\label{arcellipse_area}
\end{eqnarray}
i.e., identical to that of an ellipse with semi-axes $a$ and $b$.

In order to represent an arc, we must impose some limits on the
ArcEllipse parameters such as $b<r_c$, to ensure the arc shape, and  $a < 2\pi r_c$, to avoid closing the ArcEllipse into a ring.

\subsection{Surface brightness profile for ArcEllipses}

To create a surface brightness map with ArcEllipse isophotes, from a given radial profile, we need to simply replace the argument in a radial profile by the function
\begin{equation}
R = \sqrt{\left(\frac{ r_c \left(\theta-\theta_0 \right) }{a}\right)^2+\left(\frac{ r-r_c }{b}\right)^2} ,
\label{r_profile}
\end{equation}
which is constant over ArcEllipses. Notice that in this case the ArcEllipse must be centered at the origin (center of curvature). We can rewrite this expression as 
\begin{equation}
R = \frac{1}{b} \sqrt{[r_c \left(\theta-\theta_0 \right) (1-e) ]^2+(r-r_c)^2}=\frac{1}{b}R_{{\rm prof}},
\label{new_profile}
\end{equation}
where $e = 1 - b/a$ is the ellipticity.

\subsection{Asymmetric ArcEllipse}
\label{asymsec}
Since the gravitational arcs in general are not symmetric, we can also add some asymmetry to the ArcEllipse model. The simplest way to implement that is to consider two different values for the semi-major axis, $a_1$ and $a_2$, preserving the semi-minor axis as $b$; in other words, two ArcEllipses with different $a$ matched at  $\theta_0$.  Points whose angle $\theta$ are larger that $\theta_0$ will therefore have ellipticity
\begin{equation}
e_1 = 1 - \frac{b}{a_1},
\label{e1}
\end{equation}
whereas points whose angle $\theta$ are smaller that $\theta_0$ will have
\begin{equation}
e_2 = 1 - \frac{b}{a_2}.
\label{e2}
\end{equation}

Now the length is given by 
\begin{equation}
L=a_1+a_2= b \left( \frac{1}{1-e_1} + \frac{1}{1-e_2} \right)
\label{Lassym}
\end{equation}
and Eq. (\ref{arcellipse_area}) still holds.

In the right panel of Fig. \ref{fig_arcellipse} we show an example of an asymmetric ArcEllipse. It has the same values of $r_c$ and $\theta_0$ as the symmetric arc shown on the left panel. But now we set $e_1 = 0.9$ and $e_2 = 0.75$.

\section{PaintArcs: an implementation of ArcEllipse}
\label{painting}

PaintArcs is a code written in Python that implements a model to simulate objects with arc morphology and adds them to astronomical images. It performs the digitalization of the surface brightness profile whose radial argument is given by the Eq. (\ref{new_profile}). It also performs point spread function (PSF) convolution, to add the effect of the atmosphere and instrument, and adds the effect of Poisson fluctuations on the image counts. The code makes use of 
functions from SLtools\footnote{The SLtools library is available at http://che.cbpf.br/sltools/}, a library for image processing, catalog manipulation, 
and strong-lensing applications.

PaintArcs is currently set to use the ArcEllipse model to mimic gravitational
arc shapes. But it is built in a modular way that can accommodate other 
prescriptions. 
The surface brightness distribution currently uses the \citet{Sersic_1968} model, for which the intensity profile is given by
\begin{equation}
I_S(r)=I_e \exp \left\{-b_n\left[\left(\frac{r}{r_e} \right)^{1/n} - 1 \right]\right\},
\label{sersic}
\end{equation}
where $I_e$ is the intensity at the effective radius $r_e$ that encloses half of the total light from the model. The constant $b_n$ is defined in terms of the parameter $n$, which describes the shape of the profile. The value of $b_n$ can be computed numerically or obtained using analytical expressions that approximate its value. In this work we use the approximation of \citet{1999A&A...352..447C}, given by
\begin{eqnarray*}
b_n(n) &\approx& 2n-\frac{1}{3}+\frac{4}{405n}+\frac{46}{25515n^2}+\frac{131}{1148175n^3}- \\ &-&\frac{2194697}{30690717750n^4}+O(n^{-5}),
\label{bn}
\end{eqnarray*}
which is better than one part in $\approx 10^{-4}$ for $n>0.36$. Note that if $r_e$ is defined in a different way, the expression for $b_n$ will be different.

For an arc created with ArcEllipse we choose $b=r_e$. In other words, we distort the circular profile
 only in the tangential direction, preserving its effective size ($r_e$) in the radial direction. In this case, the ArcEllipse with S\'ersic profile is given by
\begin{eqnarray}
I(r,\theta)&=&I_S\left(R_{{\rm prof}}(r,\theta)\right)=\\
&=&
I_0\exp \left\{-b_n\left[\frac{\sqrt{[r_c \left(\theta-\theta_0 \right) (1-e) ]^2+(r-r_c)^2}}{r_e} \right]^{1/n} \right\},
\nonumber 
\label{sersic_arcellipse}
\end{eqnarray}
where $I_0 =I_e {\rm e}^{b_n}$. 

We need to normalize the intensity of the ArcEllipse $I(r,\theta)$ such that the total signal of the arc, given by
\begin{equation}
{\cal L}= \int_0^{2\pi}\int_0^{\infty} I(r',\theta')r'dr'd\theta',
\label{integral_luminosidade}
\end{equation}
is equal to the total signal of the arc in the image, which is given by
\begin{equation}
S_{total}= 10^{-0.4 (m - m_{zpt})},
\label{signal_total}
\end{equation}
where $m$ and $m_{zpt}$ are the total magnitude of the arc to 
be simulated and the zero point of the magnitude scale of the image. 
Therefore, the constant $I_0$ can be found as
\begin{equation}
I_0= \frac{10^{-0.4 (m - m_{zpt})}}{\int_0^{2\pi}\int_0^{\infty}\exp \left\{-b_n\left[\frac{\sqrt{[r_c \left(\theta'-\theta_0 \right) (1-e) ]^2+(r'-r_c)^2}}{r_e} \right]^{1/n} \right\} r'dr'd\theta'}.
\label{integral_I_0}
\end{equation}

For asymmetric ArcEllipses, the denominator of 
Eq. \ref{integral_I_0} should be replaced by a sum of two integrals, associated to the two values of ellipticity ($e_1$ and $e_2$, see \S \ref{asymsec}), 
each one on either side of $\theta_0$.

\begin{figure*}
\begin{center}
\begin{minipage}[b]{1.0\linewidth}
\includegraphics[scale=0.50]{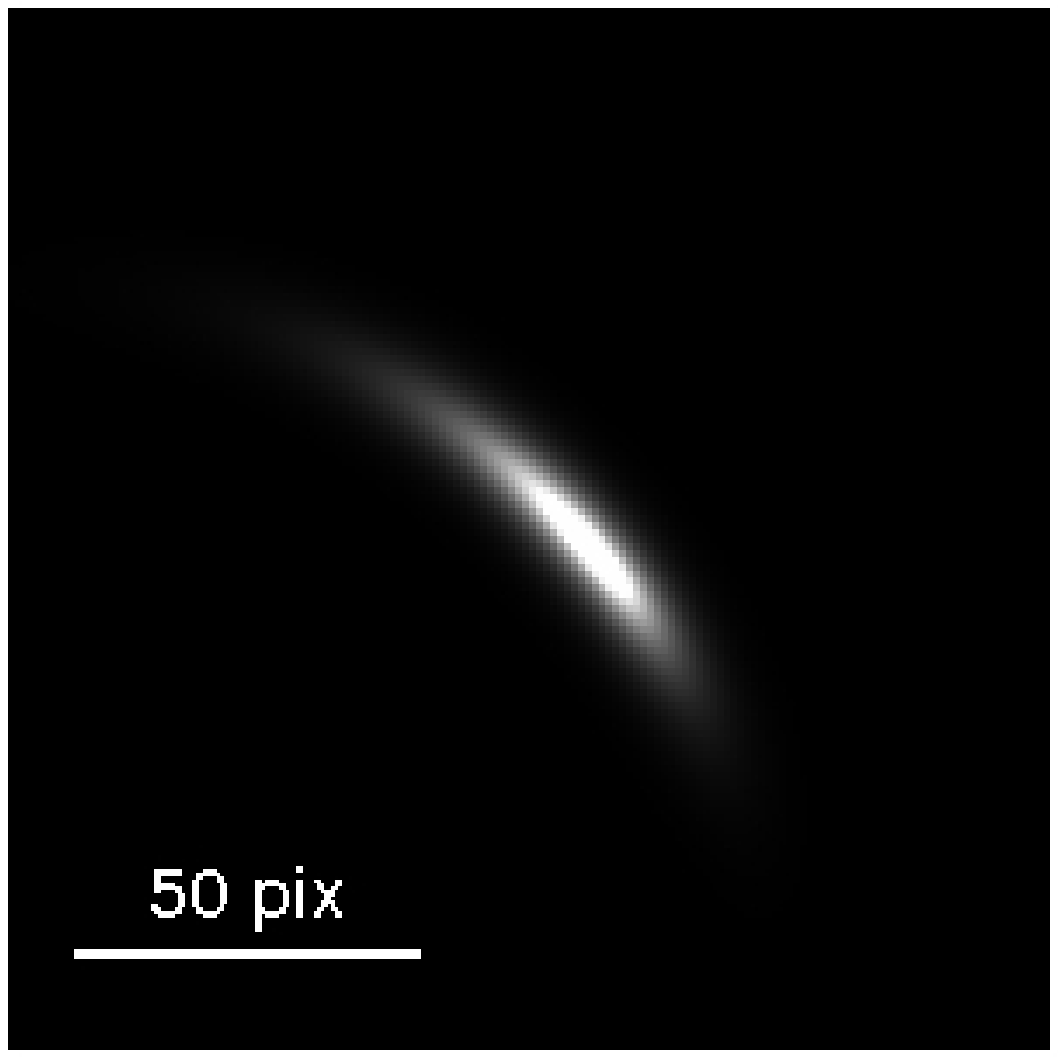}
\includegraphics[scale=0.50]{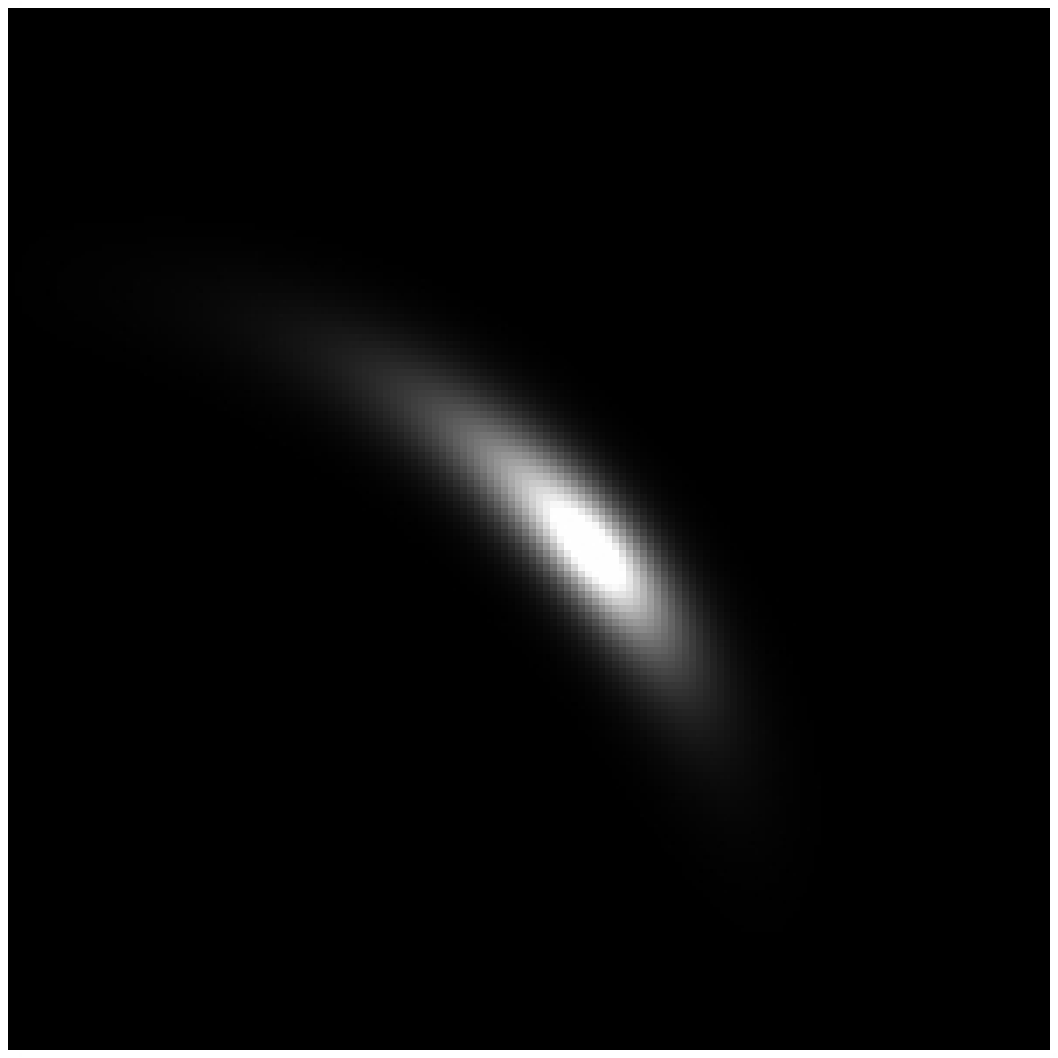}
\includegraphics[scale=0.50]{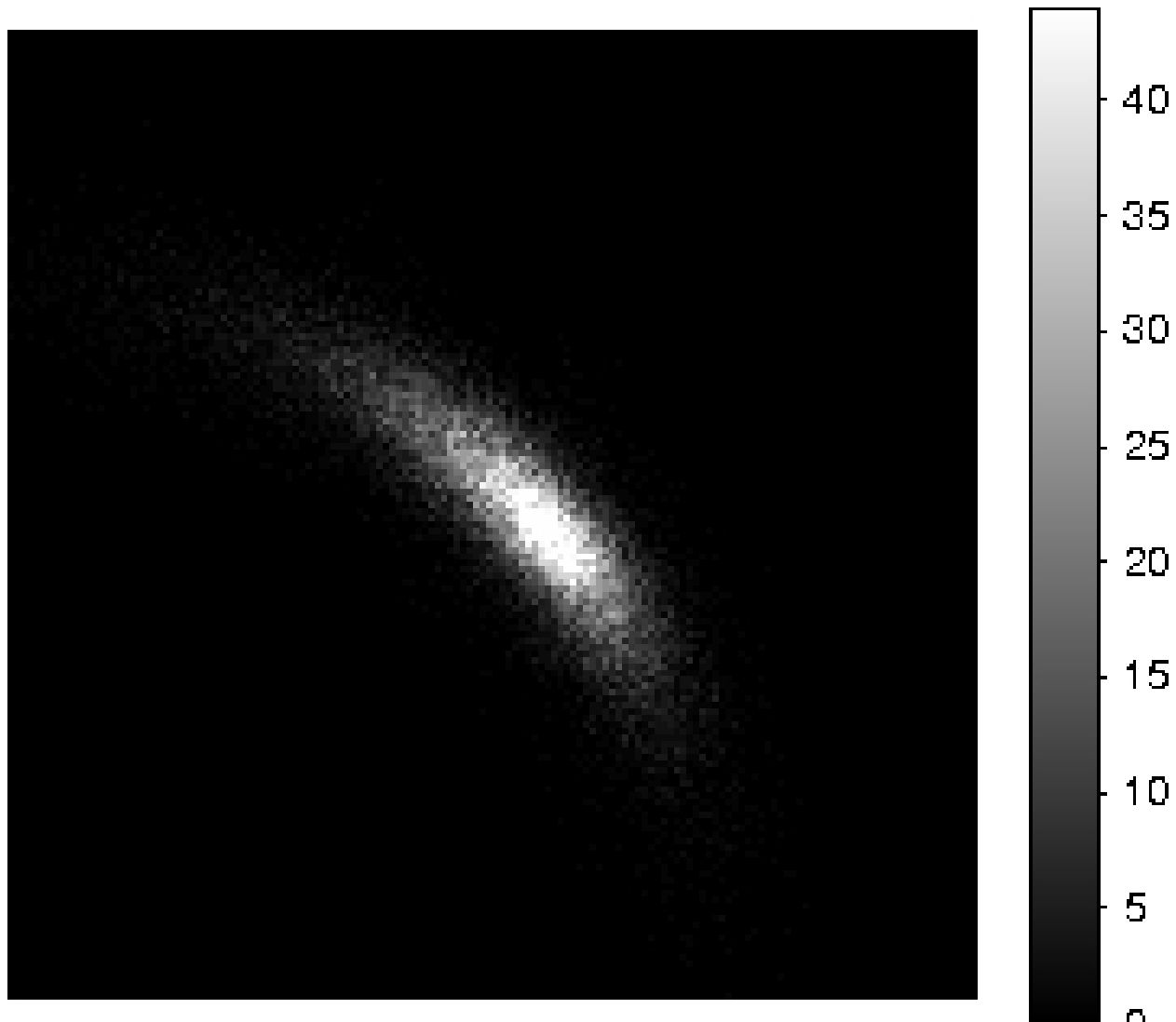}
\caption{Example of an ArcEllipse created by PaintArcs using a S\'ersic
profile. The left panel shows the pure arc, created using $r_c=17''$, $r_e=0.6''$, $\theta_0=40^o$, $n=1$, $e_1=0.85$, $e_2=0.75$, $m=21$, and $m_{zpt}=31.83$. The pixel scale of the image is $0.154''$. The middle panel shows the application 
of a Gaussian convolution (FWHM = $1.1''$) on the pure arc image. The 
right panel shows the effect of adding Poisson noise to the convolved arc.}
\label{exemplo_paint_arcs}
\end{minipage}
\end{center}
\end{figure*}

To illustrate the method presented above, we show an arc created following the process described above in the left panel of Fig. \ref{exemplo_paint_arcs}.

\subsection{Pixelation} 

To create a digitized image from a continuous intensity model, we need to 
integrate the object signal over a discrete number of pixels.
We thus need to control the
change in signal caused by the pixelation and try to keep it 
lower than some tolerance value, $\epsilon$. In a typical 
position in our image, close to the S\'ersic effective radius $r_e$, 
the profile intensity is expected to vary by

\begin{equation}
dI = \frac{\partial I}{\partial r}\vline_{r=r_c + r_e,\theta=\theta_0}\delta,
\label{di}
\end{equation}
where $\delta$ is the linear size of the pixel. Therefore, the 
fractional error in the pixel intensity is
\begin{equation}
\frac{dI}{I}=\frac{\delta}{I_{(r=r_c +r_e,\theta=\theta_0)}}\frac{\partial I}{\partial r}\vline_{r=r_c +r_e,\theta=\theta_0}.
\label{error}
\end{equation}

We keep this error within a tolerance limit by estimating the intensity at
a number of subpixels. We can find the number of subpixels $n_p$ by imposing 
$dI/I=\epsilon$ as follows
\begin{equation}
n_p = \frac{1}{\delta} = \frac{1}{\epsilon I_{(r=r_c +r_e,\theta=\theta_0)} }\frac{\partial I}{\partial r}\vline_{r=r_c +r_e,\theta=\theta_0}.
\label{np}
\end{equation}
The number $n_p$ represents the minimum integer number of subdivisions of a 
pixel required to make the error on the pixelation less than $\epsilon$.

In the example of Fig. \ref{exemplo_paint_arcs}, we used $\epsilon=0.1$ as the 
upper limit for the error on the digitalization of the arc signal. In 
this case, $n_p=4$, which means that each pixel was subdivided into 
$16$ parts. The exact value of the digitalization error, computed 
afterwards by comparing the total signal to be simulated ($S_{total}$, see
Eq. \ref{signal_total}) 
with the signal that was effectively distributed, is on the order of 
$10^{-3}$. This value is much lower than the adopted $\epsilon$, 
attesting that the 
criterium given by Eq. (\ref{np}) is based on a profile region 
where the variation in the arc intensity is stronger than average,
while the exact value of the error is computed using all pixels.
Therefore the actual pixelation error is much smaller than $\epsilon$.

\subsection{Adding the seeing effect}

To add the seeing effect to a simulated image created with PaintArcs, 
we convolve it with a 2D normalized Gaussian PSF, given by
\begin{equation}
PSF = g(x,y)= \frac{\exp \left\{-\left[\frac{x^2}{2\sigma_x^2}+\frac{y^2}{2\sigma_y^2}\right]\right\}}{\int{\int{ \exp \left\{-\left[\frac{x^2}{2\sigma_x^2}+\frac{y^2}{2\sigma_y^2}\right]\right\}dx dy}}},
\label{gaussiana}
\end{equation}
where $\sigma_x$ and $\sigma_y$ are the standard deviation in x and y 
directions. The option $\sigma_x\ne\sigma_y$ allows 
an elliptical Gaussian.

Assuming a circular Gaussian ($\sigma = \sigma_x =\sigma_y $), the standard deviation can be obtained from the FWHM of an actual
image through
\begin{equation}
FWHM = 2 \sqrt{2 \ln 2} \sigma. 
\end{equation}

We implemented a Python module to convolve images with a Gaussian kernel whose 
size is a multiple of the standard deviation ($n_{\sigma}$). 

In the middle panel of Fig. \ref{exemplo_paint_arcs} we show the result of the Gaussian convolution on the arc image previously presented (arc in the left panel of this figure), 
where we use a seeing FWHM of $1.1''$, which corresponds to $\sigma = 0.47''$, 
and $n_{\sigma} = 4$. 

Attesting the code's flexibility and modularity, new implementations 
of PaintArcs are being developed to use alternative seeing 
kernels, such as a 2D Moffat function.

\subsection{Adding Poisson noise}

In a quantum detector such as a CCD the photon counts follow a Poisson distribution, 
\begin{equation}
P(x,\mu)=\mu^x\frac{e^{-\mu}}{x!},
\label{poisson}
\end{equation}
where $\mu$ is the mean of the distribution. 

We add Poisson noise to an ArcEllipse produced by PaintArcs by randomly
picking values from the distribution given by equation (\ref{poisson}) with the 
mean equal to the signal in each pixel of the simulated image after convolution
with the PSF. This is done with a separate module that generates 
random numbers from a Poisson distribution. 

In the right panel of Fig. \ref{exemplo_paint_arcs} we show the result of adding Poisson noise to the arc of the middle panel of the same figure. 

\subsection{Applying PaintArcs}

After the pixelation 
of the surface brightness profile of the arc created with PaintArcs with a S\'ersic profile and adding seeing and Poisson noise to its image, 
PaintArcs can paste this simulated arc onto a real image (for example one containing a galaxy cluster),
which can be specified in its input configuration file. In this case, 
PaintArcs adds the arc model parameters to the image header for bookkeeping
purposes.

\begin{figure}[!htb]
\begin{center}
\resizebox{8cm}{!}{\includegraphics{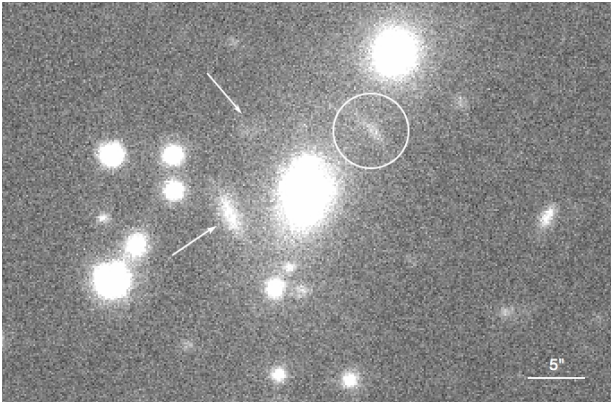}}
\caption{Example of the application of PaintArcs on a real image: the circled arc was created by the code using an ArcEllipse+S\'ersic model with $m = 21.5, e_1= 0.9, 
e_2 = 0.8, r_c = 7.7'', \theta_0 = 40^o, n = 3$, $r_e = 0.8'', FWHM = 0.7''$ 
and was added to the image of the galaxy cluster SOGRAS0328+0044, where two real arc candidates (indicated by arrows) were found.} 
\label{result-sogras}
\end{center}
\end{figure}

Figure \ref{result-sogras} shows an example of an arc simulated with PaintArcs.
It was added to the image of the galaxy cluster SOGRAS0328+0044, which was 
imaged by the SOAR Gravitational Arc Survey \citep[SOGRAS,][]{sogras}.Two real arc candidates are also found around the cluster's 
brightest galaxy (indicated by arrows), and may serve as a visual guideline for the accuracy of the PaintArcs
outputs. The simulated ArcEllipse is circled in the figure and 
was generated with the parameters $m = 21.5, e_1= 0.9, e_2 = 0.8, 
r_c = 7.7'', \theta_0 = 40^o, n = 3$,  and $r_e = 0.8''$. It was convolved 
with a Gaussian PSF whose FWHM was obtained from the image ($FWHM = 0.7''$). 
The origin of the coordinate system of the arc coincides with the center of the cluster's brightest galaxy. The pixel size of this image is $0.154''$. 

\section{ArcFitting}
\label{arcfitting}

Motivated by the need for tools for measuring arc parameters, such as its length, 
width, curvature, and surface brightness, we also developed a tool that fits
these properties to real (or realistically simulated) gravitational arcs. 
We named it ArcFitting. 

Currently, ArcFitting is based on the ArcEllipse model with a S\'ersic profile. Therefore, it performs 
measurements on actual arc images that lead to estimates of the ArcEllipse 
parameters. Similarly to PaintArcs, ArcFitting is also a Python-based code 
that was designed in a modular way to allow different shape- and profile 
models to be adopted in the future.

In Section \ref{arcellipse} we showed that the ArcEllipse depends on the following parameters (see Eq. \ref{new_profile})
\begin{itemize}
\item $x_0$: $x$ position of the point where the arc is centered (pixel)
\item $y_0$: $y$ position of the point where the arc is centered (pixel)
\item $r_c$: distance of the arc center relative to $(x_0,y_0)$ (pixel)
\item $\theta_0$: orientation of the arc center relative to the $x$ axis (radians)
\item $e_1$ and $e_2$: ellipticities of the arc
\item $b$: minor semi-axes of the arc (pixel).
\end{itemize}
If the surface brightness distribution of the arc follows a S\'ersic profile, we add three more parameters to the fitting:
\begin{itemize}
\item $r_e$: radius that encloses half of the total luminosity of the S\'ersic profile (pixel)
\item $n$: index that describes the shape of the S\'ersic profile 
\item $I_0$: intensity of the S\'ersic profile at the center of the arc (r = 0).
\end{itemize}

If we choose $b = r_e$, as we do in PaintArcs, we obtain nine fitting 
parameters. Therefore, the set of parameters that represents the ArcEllipse with a S\'ersic profile is $\mbox{\boldmath${p}$} = 
(x_0,y_0,r_c,\theta_0,e_1,e_2,r_e,n,I_0)$.

\subsection{Creating an arc image to be fitted}
\label{create_arc_image}

The ArcFitting code works on a {\it postage stamp} image of an observed arc (i.e., an image where the pixels not belonging to the arc are set to zero), 
which we call the {\it arc image}. To create the arc image we first 
run SExtractor 2.8.6 \citep{1996A&AS..117..393B} 
on the original image that contains the arc. SExtractor then creates a {\it segmentation}
image, which assigns a single value (corresponding to the sequential number
of the object in the output image catalog) to all pixels 
that it associates to the object. 
The arc image is then created by 
copying all arc pixels defined by the segmentation image, but using their 
values as in the original image. All other pixels in the arc image 
are assigned a zero value. The
size of the arc image in the x and y directions are multiples
of the arc's major and minor axis values as found by SExtractor. 
The procedure described above to create the arc image makes use of 
functions from SLtools.

\subsection{Parameter determination from the arc image}
\label{arcpars}

In the following we describe the measurements performed by ArcFitting to 
find the ArcEllipse parameters. But the ArcFitting is not only 
restricted to the model parameters, which currently are the nine ArcEllipse+S\'ersic 
parameters listed above. It also attempts to measure
other commonly used arc parameters, such as its length $L$ and width $W$.

\subsubsection{Determining the ArcEllipse parameters}
\label{ArcEllipseparam}

First, ArcFitting finds the two farthest points of the arc image, 
$p_1$ and $p_2$. For that, it uses any reference point to find the 
first extreme as the farthest point from the reference one. It then finds 
the farthest point from this first extreme through the farthest-of-farthest 
method. The code then names the extreme point whose angle is larger 
(smaller) than $\theta_0$ (see Eq. \ref{theta0} below) as $p_1$ ($p_2$).

After that, ArcFitting obtains the points along the bisectrix 
between $p_1$ and $p_2$. This line is perpendicular to the line segment 
that connects these two extreme points.
It then finds $p_3$, which is the mean point of the arc along the $b$ 
direction.

This process is repeated twice to obtain the points $p_4$ and $p_5$. $p_4$ is 
the mean arc position along the bisectrix line between $p_1$ and $p_3$,
whereas $p_5$ is the mean arc position along the bisectrix line between $p_2$ 
and $p_3$. This method is an application of the Mediatrix method \citep{mediatrix} for $n_b=2$, where $n_b$ is the number of bisections. 
The length ($L$) of the arc is then computed as the sum of the 
line segments connecting the five points

\begin{equation}
L = \overline{p_1 p_4} + \overline{p_4 p_3} + \overline{p_3 p_5} + \overline{p_5 p_2}.
\label{length}
\end{equation}

\begin{figure*}[!htb]
\centering
\begin{minipage}[h]{0.95\linewidth}
\includegraphics[scale=0.35]{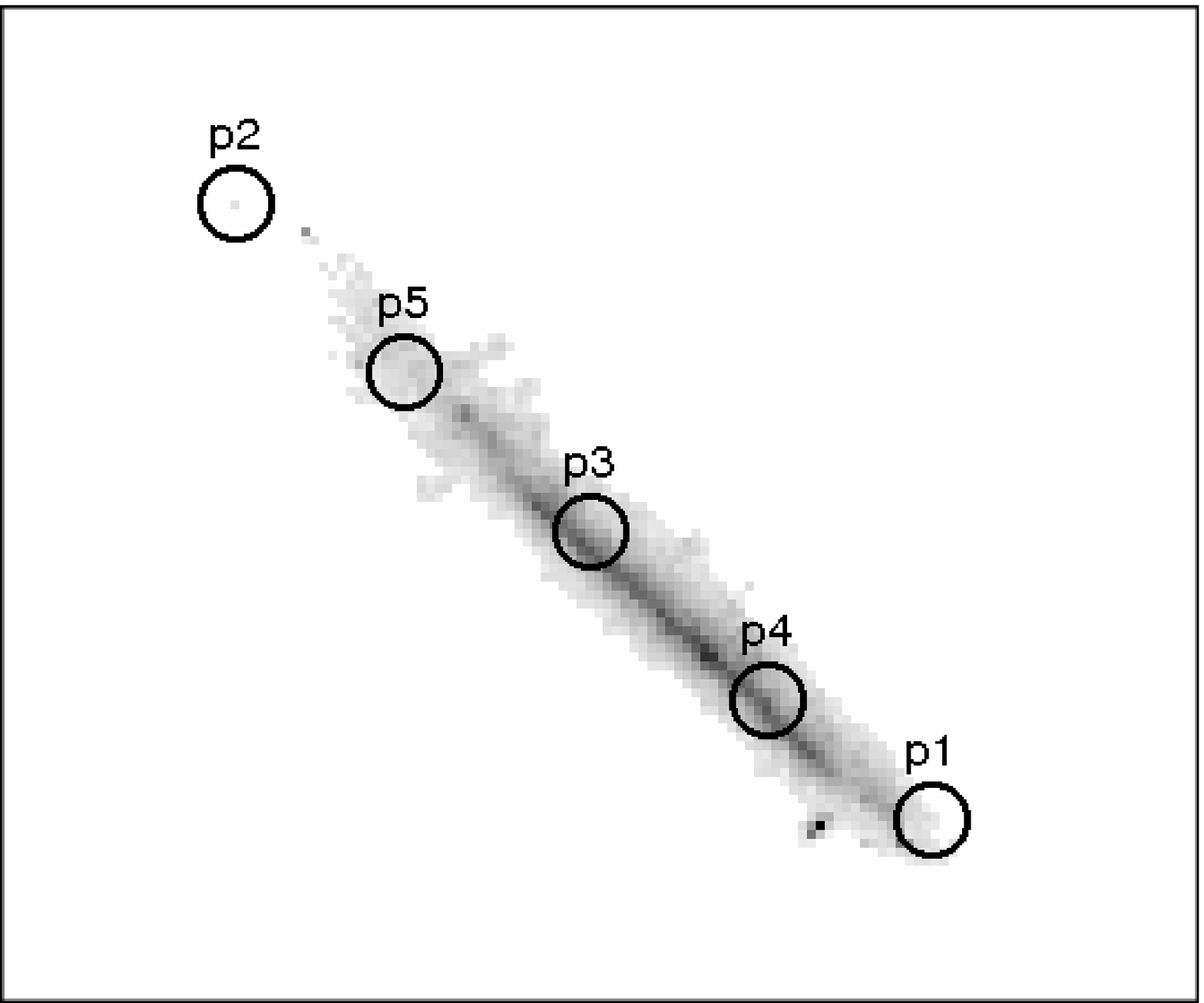}\hspace{1cm}
\includegraphics[scale=0.35]{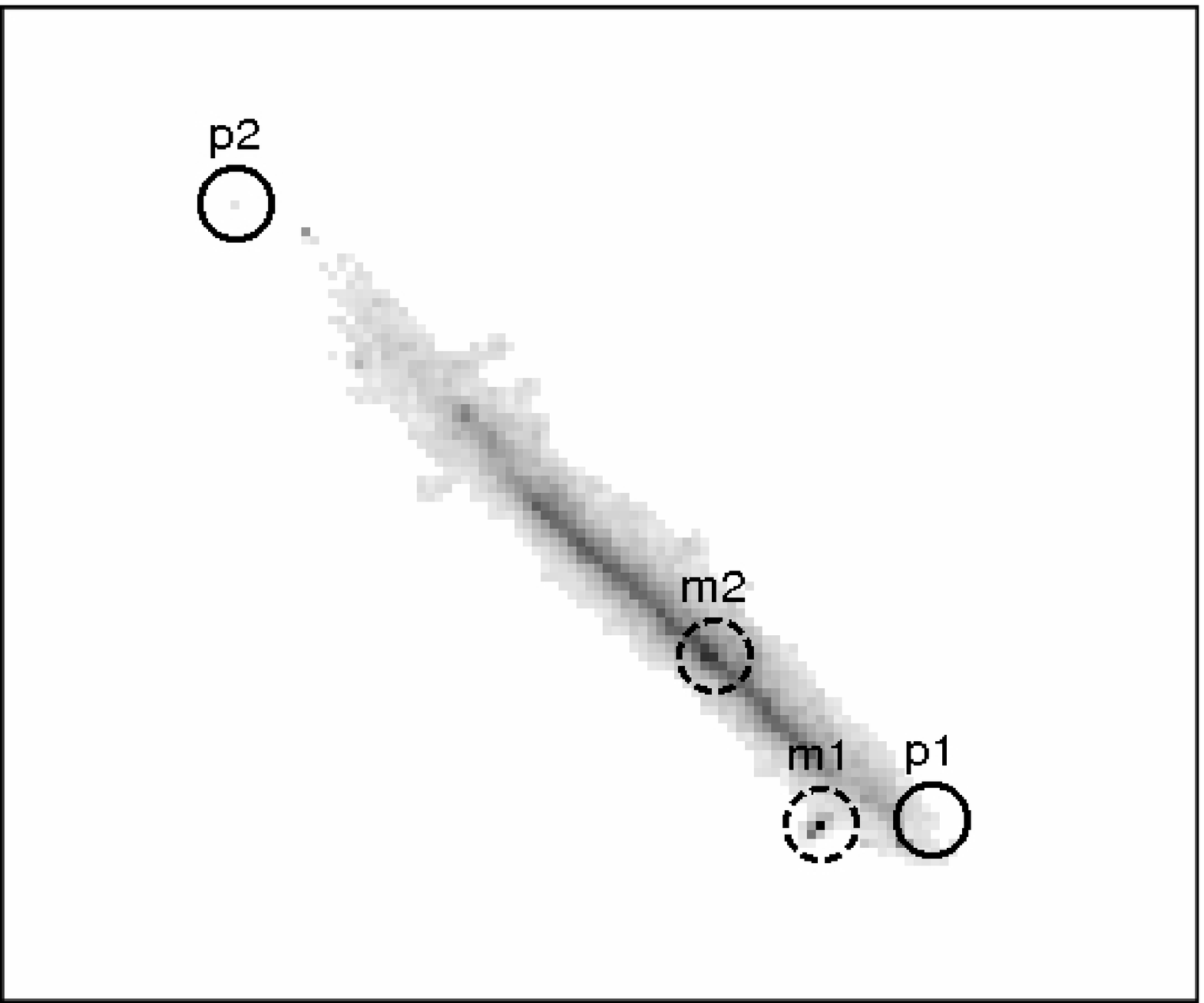}
\caption{Illustration of the process of tracing a real arc in its image. The left panel shows the five $p$ points used to compute the arc length. The right panel shows the high-intensity $m$ points (dashed circles) used to find $p_{max}$ for the same arc. The center and radius of curvature are determined as the circle containing the points $p_1, p_2$, and $p_{max} = m_2$. The width is based on the
line through the central point $p_3$, whereas the asymmetric major-axes $a_1$ and $a_2$
are the lengths of the segments connecting $m_2 = p_{max}$ to the arc extremes $p_1$ and $p_2$.}
\label{paints_arc}
\end{minipage}
\end{figure*}

In the left panel of Fig. \ref{paints_arc}, we illustrate this process of tracing the arc image and determining its length. The five points $p_i, i=1,5$ are shown along the arc image. The image used as an example in the figure is taken from an HST arc sample that will be presented in \S \ref{hst_arcs}. 

The ArcEllipse parameters $x_0$, $y_0$ and $r_c$ are found by 
determining the unique circle that passes through the extreme points, 
$p_1$ and $p_2$, and through the point with maximum intensity, $p_{max}$.
The latter point is found as the arc point with maximum intensity among those that lie within $0.5 \sigma_b$ from the line 
segments of Eq. \ref{length}, where $\sigma_b$ is the dispersion of the points along the bisectrix line containing $p_3$. This procedure is intended to eliminate 
arc image pixels located far away from the traced arc, which may be 
associated to a cosmic ray, or to a residual from the process
of object segmentation and deblending, as these pixels 
would bias the $p_{max}$ estimate. The determination of $p_{max}$ is depicted in the right panel of Fig. \ref{paints_arc}. The pixel of maximum intensity in the entire image, $m_1$, failed to satisfy
the $0.5 \sigma_b$ distance criterion, in this case measured from the
$\overline{p_1 p_4}$ segment. The second maximum, $m_2$, however, is well-placed
along the observed arc. 

The width of the arc is taken as $W = 4 \sigma_b$. 
The ArcEllipse semi-minor axis $b$ is then taken as half of the width 
($b = W/2 = 2 \sigma_b$).

The angle $\theta_0$ is obtained using the expression
\begin{equation}
\theta_0 = \arctan \left(\frac{(y_c - y_0)}{(x_c - x_0)}\right),
\label{theta0}
\end{equation}
where $(x_c,y_c)$ are the coordinates of the $p_{max}$ point. 

The asymmetric ArcEllipse ellipticities $e_1$ and $e_2$ are obtained using Eqs. \ref{e1} and \ref{e2}, respectively, where $a_1$ is the size of the line segment $\overline{p_1 p_{max}}$ and $a_2$ is the size of the line segment $\overline{p_{max} p_2}$.

The procedure outlined above is different compared to previous
works. Some studies, particularly of observed arcs, quote $L$, 
$L/W$, or other arc parameters, without clearly mentioning how they 
are determined. Other sources 
use no more than three points to infer $L$, either using a sum of arc segments, 
as in eq. \ref{length}, or fitting a circular arc through 
them \citep{Oguri2002, Dalal2004, Li2005, 2005ApJ...627...32S, 2005ApJ...633..768H, 
2008A&A...482..403M, Horesh2010, 2012ApJ...749...38M}. \citet{2004A&A...416..391L} and 
\citet{Kausch2010} measured momenta over arc images to derive their sizes.

Width measurements are usually based
on the area of the arc divided by its length assuming some arc geometry,
usually a rectangle. \citet{2008A&A...482..403M}, however, measured widths at
several different positions along the arc length and assumed a median value.
Some authors also account for seeing effects when measuring $W$ \citep{2008A&A...482..403M, Zaritsky2003}.

Arc orientation and curvature are not as common to find. One such
reference is \citet{Luppino1999}, who used the chord length from the
arc extremes and its sagittal depth to estimate $r_c$. Estimates of
$\theta_0$ can also be found in \citet{Oguri2002} and \citet{Miralda1993}.

We are unaware of previous studies that accommodate a possible asymmetry in the
arc's shape. Finally, another element lacking in previous attempts to
measure arc properties is the surface brightness distribution. This is the
topic of the next subsection.

\subsubsection{Determining the S\'ersic parameters}

We use $\chi^2$ minimization to determine the S\'ersic profile parameters, $I_0$, $n$ and $r_e$. This $\chi^2$ minimization works as follows.

We consider a data set that represents the information of $N$ pixels in an arc 
image. We denote $I_{img}(j)$ as the count of the $j^{th}$ pixel of the arc 
image and $\sigma^2_{j}$ as its associated uncertainty, 
where $j = (1,N)$. We also denote $I_{arc}(j,\mbox{\boldmath${p}$)}$ as the 
theoretical count of the $j^{th}$ pixel from ArcEllipse with a 
S\'ersic profile. This theoretical
intensity depends on the pixel position and on the set of parameters 
$\mbox{\boldmath${p}$}$. Except for $I_0$, $n$ and $r_e$, whose values 
are unknown, the other parameters are fixed 
by the procedure described in Sect. \ref{ArcEllipseparam}.
Assuming that the pixel counts $I_{img}(j)$ are independent from each other, 
we must minimize the quantity
\begin{equation}
\chi^2 = \sum_{j} \frac{\left(I_{arc}(j,\mbox{\boldmath${p}$})-I_{img}(j)\right)^2}{\sigma_{j}^2}
\label{chi2}
\end{equation}
to estimate the remaining parameters that provide the best fit to 
the data.

In general, we use the reduced $\chi^2$
\begin{equation}
\chi^2_{red} = \frac{\chi^2}{N-1}
\label{chi_red}
\end{equation}
as a measure of the concordance between the model and the data. If this value 
is much higher than unity, the model is not considered a good description of the data.

We assume that 
the pixel counts follow a Poisson 
distribution, such that 
\begin{equation}
\sigma_{j}^2 = I_{img}(j),
\label{sigma_j}
\end{equation}
and Eq.(\ref{chi_red}) becomes
\begin{equation}
\chi^2_{red} = \frac{1}{N -1} \sum_{j} \frac{\left(I_{arc}(j,\mbox{\boldmath${p}$})-I_{img}(j)\right)^2}{I_{img}(j)}.
\label{chi2red2}
\end{equation}

The $\chi^2_{red}$ function minimization is carried out with Pyminuit\footnote{http://code.google.com/p/pyminuit/}, which is an extension 
for Python modules of the Minuit\footnote{http://seal.web.cern.ch/seal/snapshot/work-packages/mathlibs/minuit/} functions. 

The initial guesses for $I_0$ and $r_e$ are determined through measurements on the arc image. We consider the intensity of the point $p_{max}$ as the initial estimate for $I_0$. The semi-minor axis $b$, whose estimate 
was described above, is the initial guess for $r_e$. Given the lack of a direct method to estimate the parameter $n$ from the image, the initial guess of this parameters is found by using the algorith Scan from Pyminuit, which calculates $\chi^2_{red}$ for a range of $n$ values keeping all other parameters in the $\mbox{\boldmath${p}$}$ set fixed, i.e., the morphological parameters as determined in
the previous subsection, and $I_0$ and $r_e$ fixed at their initial guesses. 

After determining the initial guesses for $I_0$, $n$ and $r_e$, we use the Pyminuit algorithm Migrad, which is an optimized 
gradient-based minimum search algorithm, to perform the $\chi^2_{red}$ minimization and to find the best-fitting values for these parameters.

The approach outlined here to estimate the shape parameters directly from
the image and to apply a $\chi^2$ minimization to the S\'ersic profile 
proved to be the most effective. We tried other alternatives, such as 
$\chi^2$ minimization (using Migrad) of all shape- and S\'ersic
parameters, or an iterative method that combines Migrad for 
some parameters and brute force (Scan) for others. In many of the
cases we tried, these methods did not converge, or resulted in a final
$\chi^2_{red}$ much larger than unity. These failures are probably caused by 
a combination of strong degeneracies among the ArcEllipse and S\'ersic model 
parameters and by the limited degrees of freedom when considering the number 
of available pixels in the arc image.


\subsection{Testing the ArcFitting algorithm}
\label{validation}

To test and validate the ArcFitting code, we applied it to images of 
pure arcs (noiseless and with no seeing convolution) produced with PaintArcs, 
using the ArcEllipse+S\'ersic model.

As an example, we considered an arc with true parameters as shown in the middle column of Table \ref{result_paint_arcs}. The values recovered by ArcFitting are displayed in the right column, showing an excellent agreement with the input values used by PaintArcs.
This situation is typical of the validation tests. Indeed, ArcFitting was very successful in recovering the ArcEllipse and S\'ersic parameters in this idealized case.
This is expected since both PaintArcs and ArcFitting use the same analytical expression for the arc shape and surface brightness profile, but demonstrates the validity of the whole process of determining the morphological parameters and using the initial guesses for scanning and fitting the S\'ersic parameters.

\begin{table}[h]
\caption{Input and fitted ArcEllipse and S\'ersic parameters for an arc created
with PaintArcs using the ArcEllipse prescription and a S\'ersic surface
brightness profile. }
\label{result_paint_arcs}
\centering
\begin{tabular}{c c c}
\hline
\hline
Parameter     &  True value      & Recovered value \\
\hline
$x_0$ (pix)         &      100.00      &         98.50\\

$y_0 $ (pix)        &      100.00         &     96.88\\

$r_c$ (pix)        &      50.00        &    52.85 \\

$r_e$ (pix)           &     5.20        &     5.96\\

$\theta_0$ (rad)     &    0.70      &     0.72\\

$I_0 $ (counts)        &      100.00    &       102.81\\

$n $            &     1.00        &        1.06\\

$e_1 $      &       0.90     &      0.88 \\

$e_2 $      &       0.80     &      0.77 \\

\hline
\end{tabular}
\end{table}

\section{ArcFitting application to real and simulated arcs}
\label{fitapplication}

We applied the ArcFitting code to three different sets of arcs. The first one consists of realistically simulated arcs. The other two sets are of real arcs, one from ground-based images and the other from HST images.

We used the residual signal contrast $\Delta S / S$, defined as
\begin{equation}
\frac{\Delta S}{S} =\frac{ \sum\limits_{j} \left( I_{img}(j) - I_{arc}(j,\mbox{\boldmath${p}$})\right)}{\sum\limits_{j}I_{img}(j)}, 
\label{deltaS}
\end{equation}
and the reduced $\chi^2$, given by Eq. (\ref{chi2red2}), as metrics for the concordance between the ArcEllipse+S\'ersic model, resulting from
ArcFitting, and the data. Both quantities are computed using only the pixels that belong to the arc 
as described in \S \ref{create_arc_image}. 

\subsection{Applying ArcFitting on arcs simulated with AddArcs}
\label{add_arcs}

For this first application we used a set of arcs simulated with the AddArcs code (Brandt et al., in prep.), which was originally designed for Dark Energy Survey image simulations. The code uses as basic inputs a catalog of lenses, a catalog of sources, the cosmological parameters, and the properties of the images on which the arcs will be simulated (e.g., pixel scale, magnitude zero points, etc.). For the simulations used in this work, the input lens catalogs were given by the Carmen Las Damas N-body cosmological simulation\footnote{http://lss.phy.vanderbilt.edu/lasdamas/}  \citep{2011AAS...21724907M} and the lenses are modeled as a projected Navarro--Frenk--White \citep[][hereafter NFW]{nfw96,nfw97} profile with elliptical mass distribution. 
The redshift and mass of each lens are taken from the catalog, but the NFW concentration parameter is taken from the fits of \citet{Neto2007}, since the simulations do not have sufficient precision to determine it accurately. The sources are modeled as elliptical S\'ersic brightness distributions whose parameters are obtained from fits to galaxies in the Hubble Ultra-Deep Field (UDF) given by \citet{UDFmorpho}, which provides the input source catalog. This same catalog contains the photometric redshift information from multi-waveband data on the UDF.

The sources are divided into redshift bins whose spatial distribution is uniform in each bin. For each lens (i.e., dark matter halo) in the catalog, a number of sources is lensed using the Gravlens \citep{2011ascl.soft02003K} software. The number of sources to be lensed is obtained from the surface density distribution of the input source catalog. A first selection of sources is made using the local magnification computed with Gravlens and then the same code is used to produce a full brightness distribution by lensing finite S\'ersic sources from the UDF catalog. Finally, images are identified and are dimensionally estimated by SExtractor, and only elongated images are selected by the code. We selected a subset of nine arcs generated from these simulations to have a similar sample size as the two other samples.

\begin{figure*}[!htb]
\begin{center}
\begin{minipage}[h]{1.0\linewidth}
\includegraphics[scale=0.25]{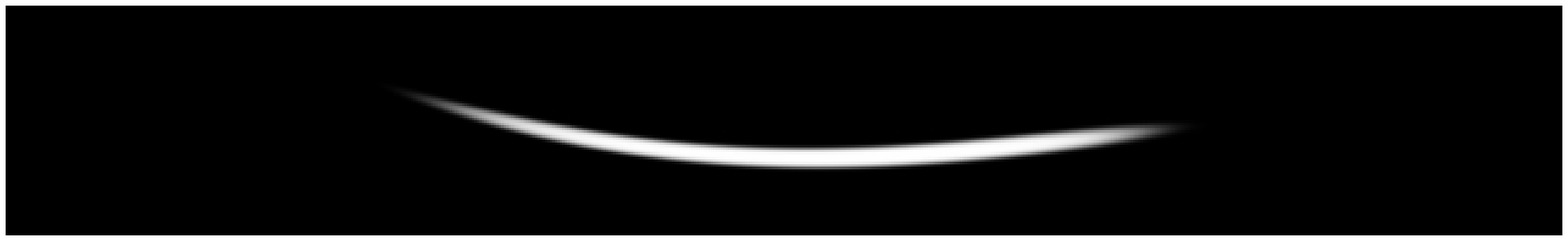}
\includegraphics[scale=0.25]{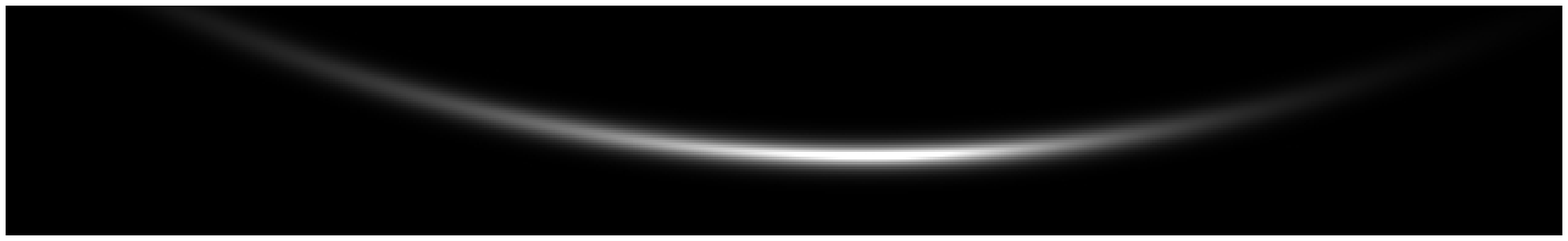}
\includegraphics[scale=0.25]{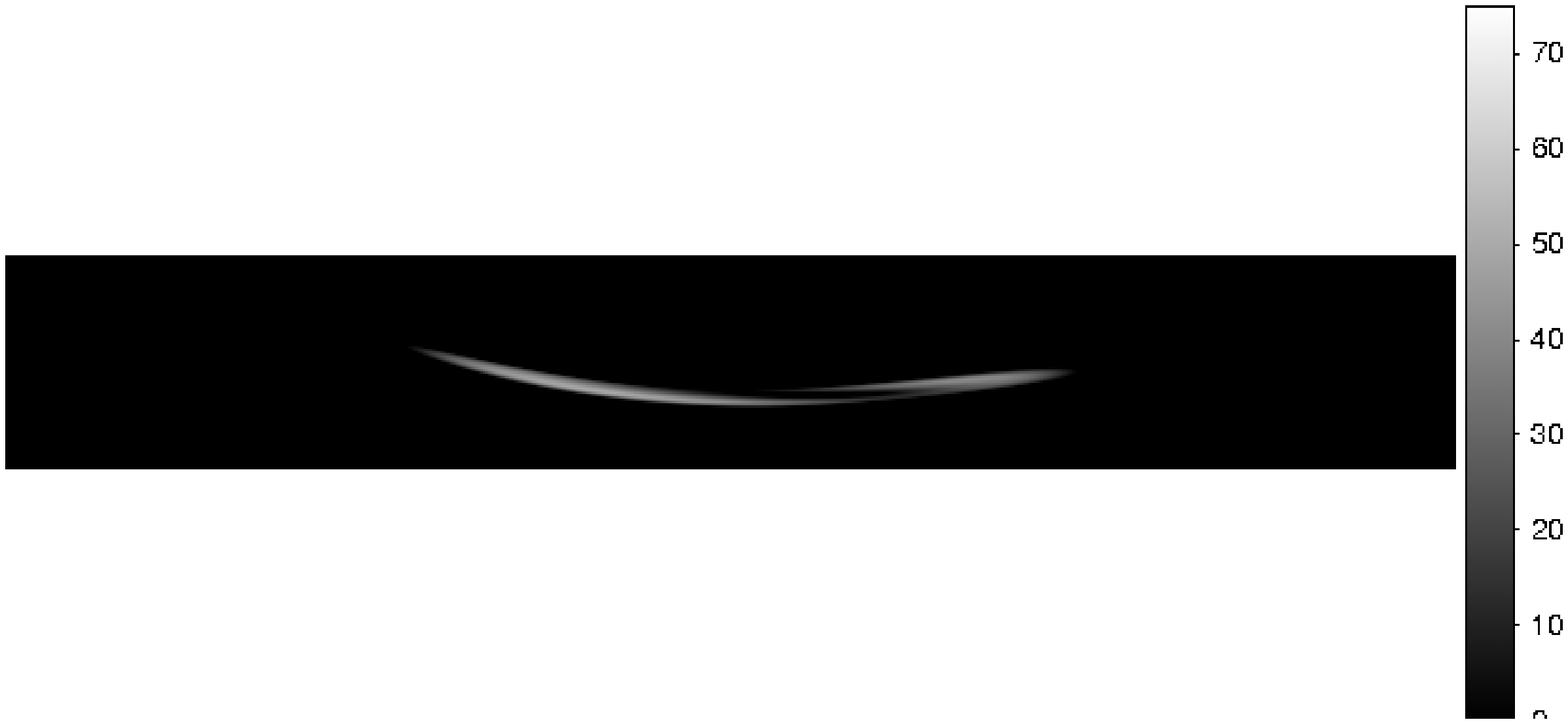}\\
\includegraphics[scale=0.25]{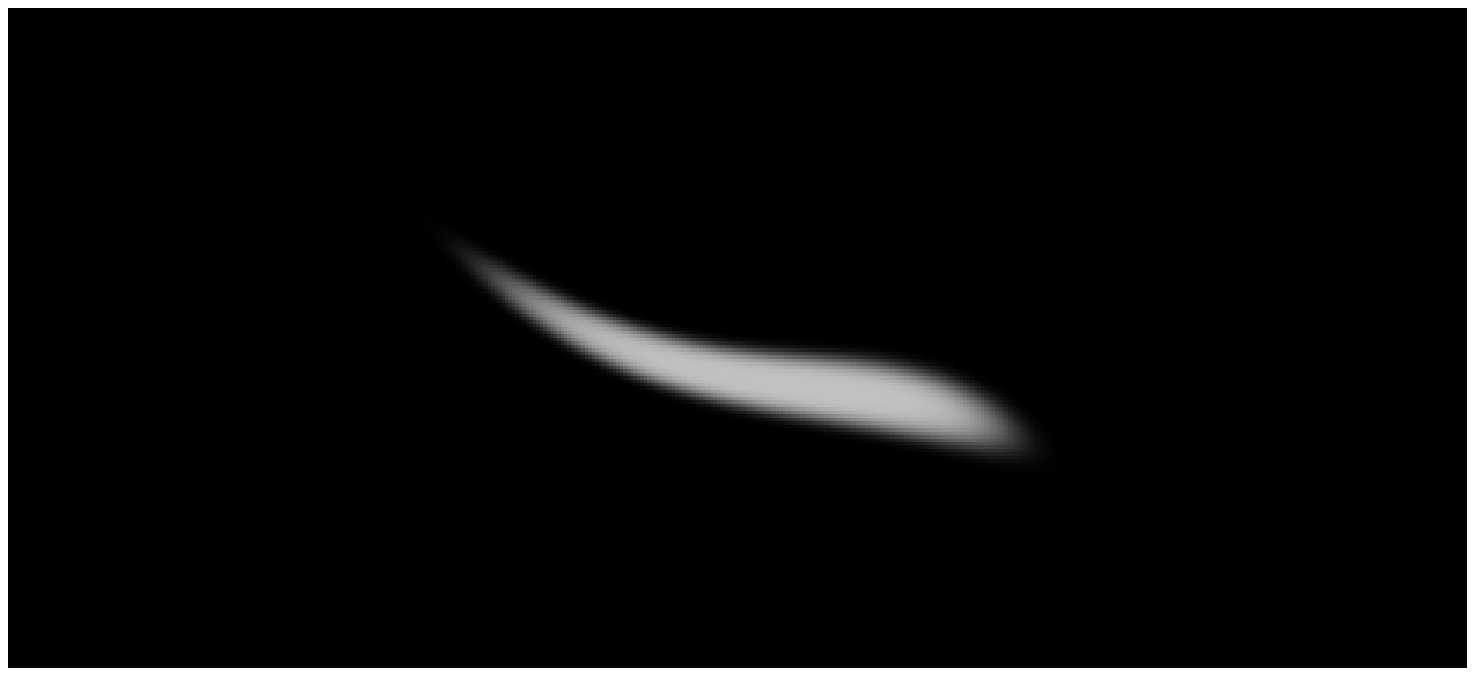} 
\includegraphics[scale=0.25]{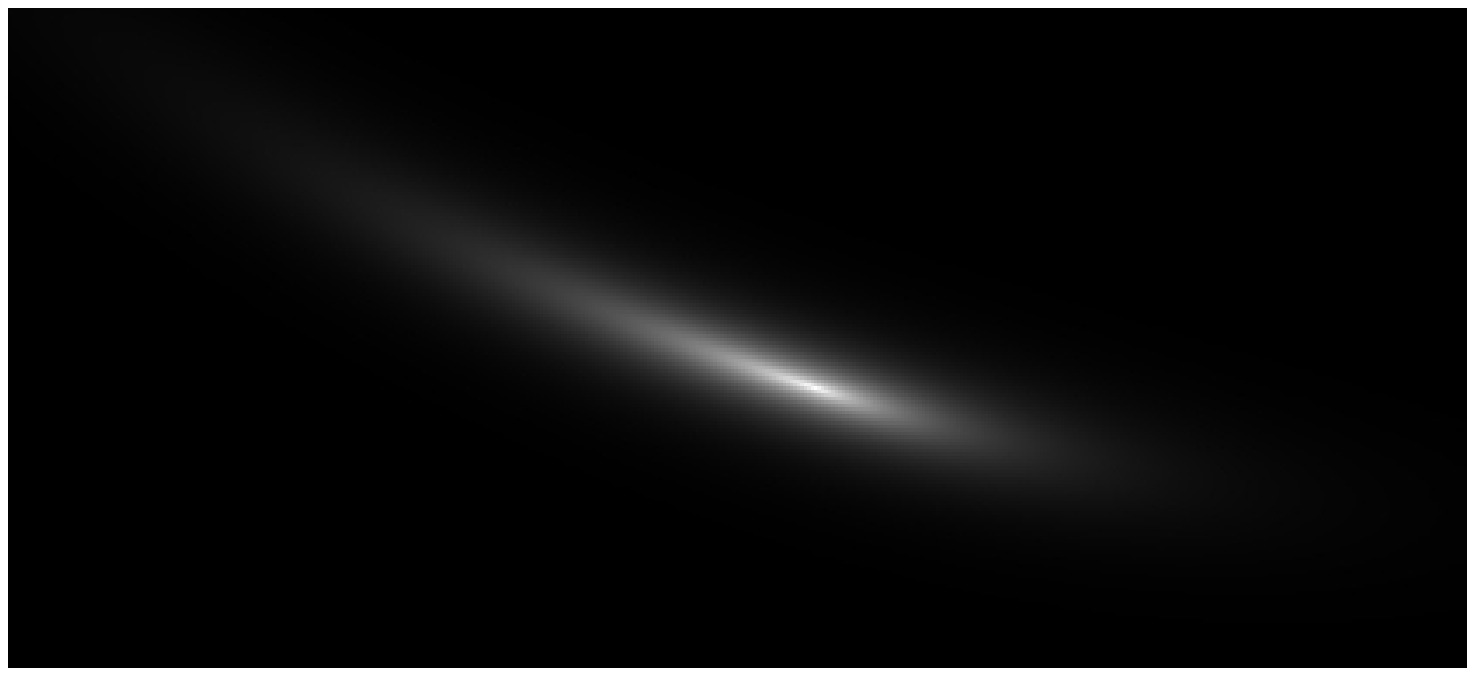}
\includegraphics[scale=0.25]{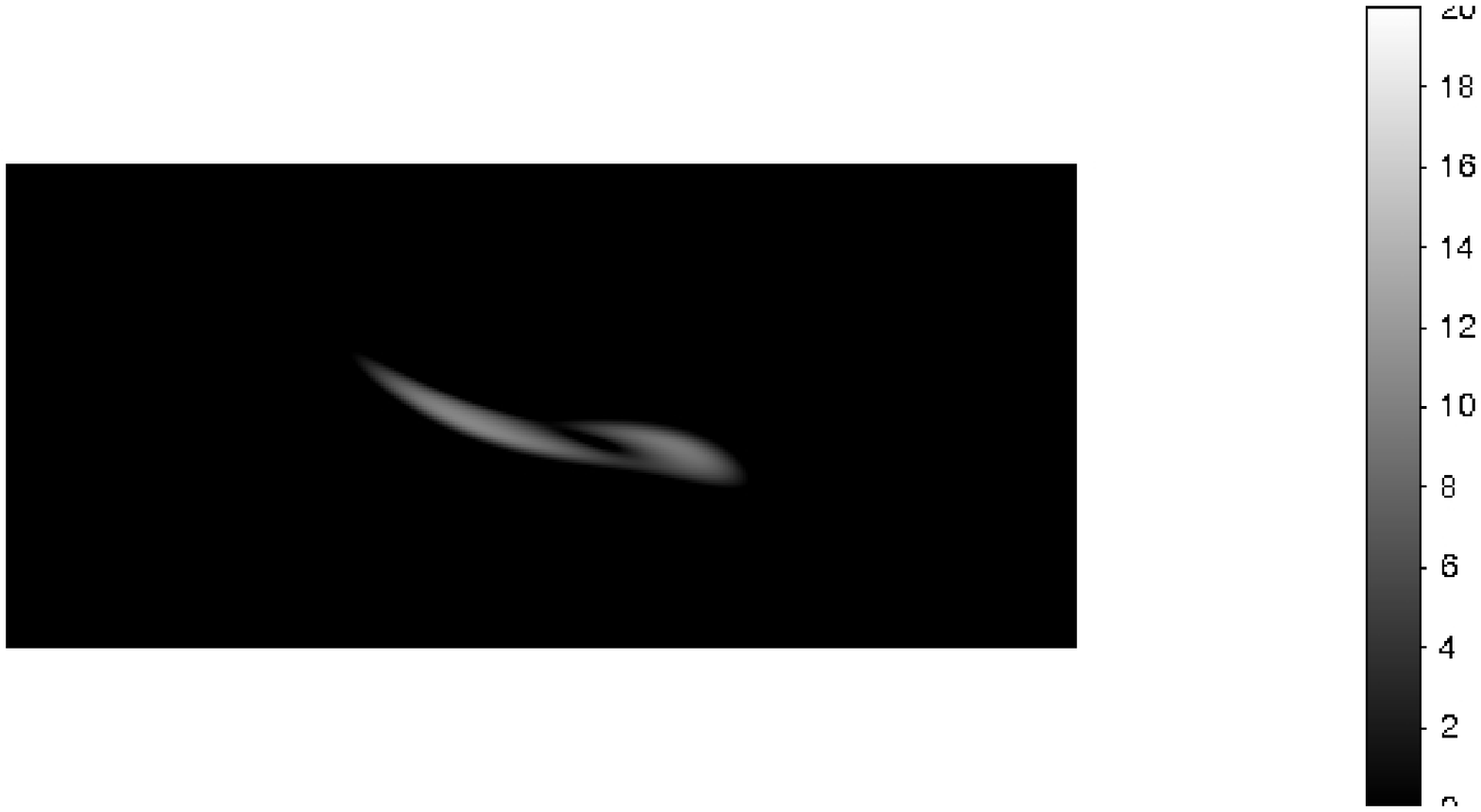}\\
\includegraphics[scale=0.25]{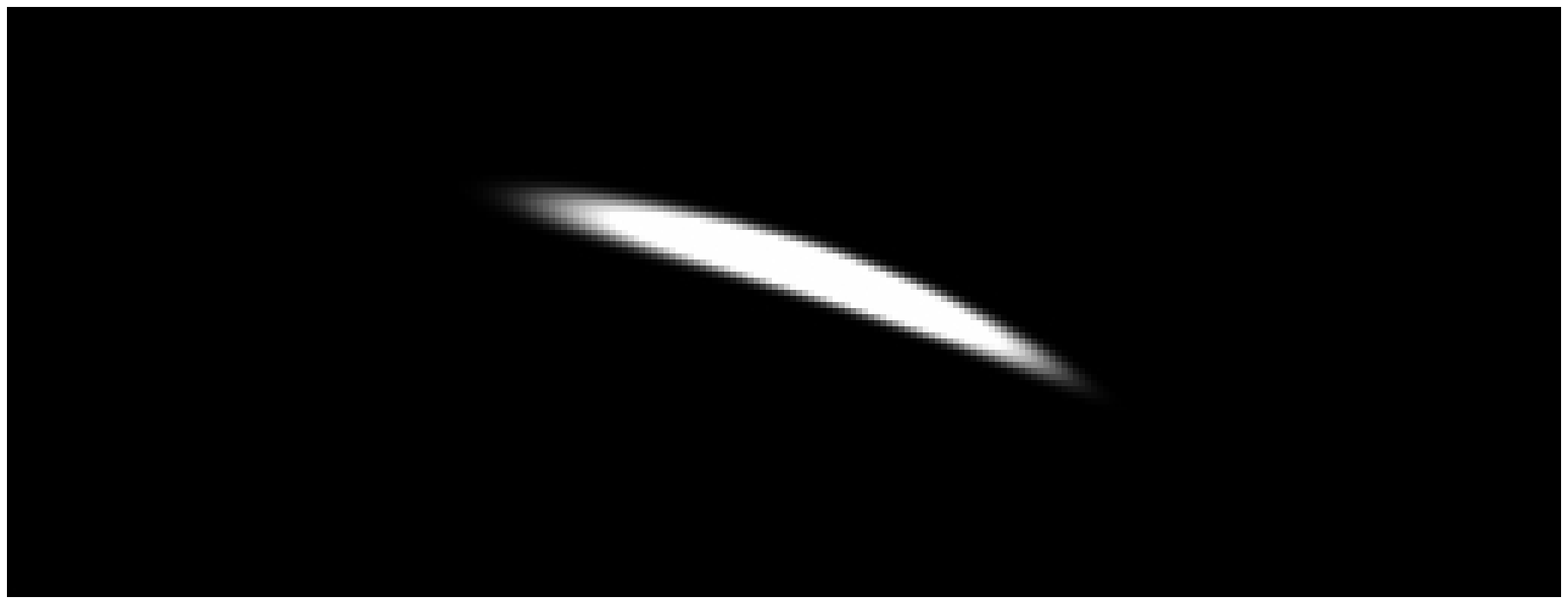} 
\includegraphics[scale=0.25]{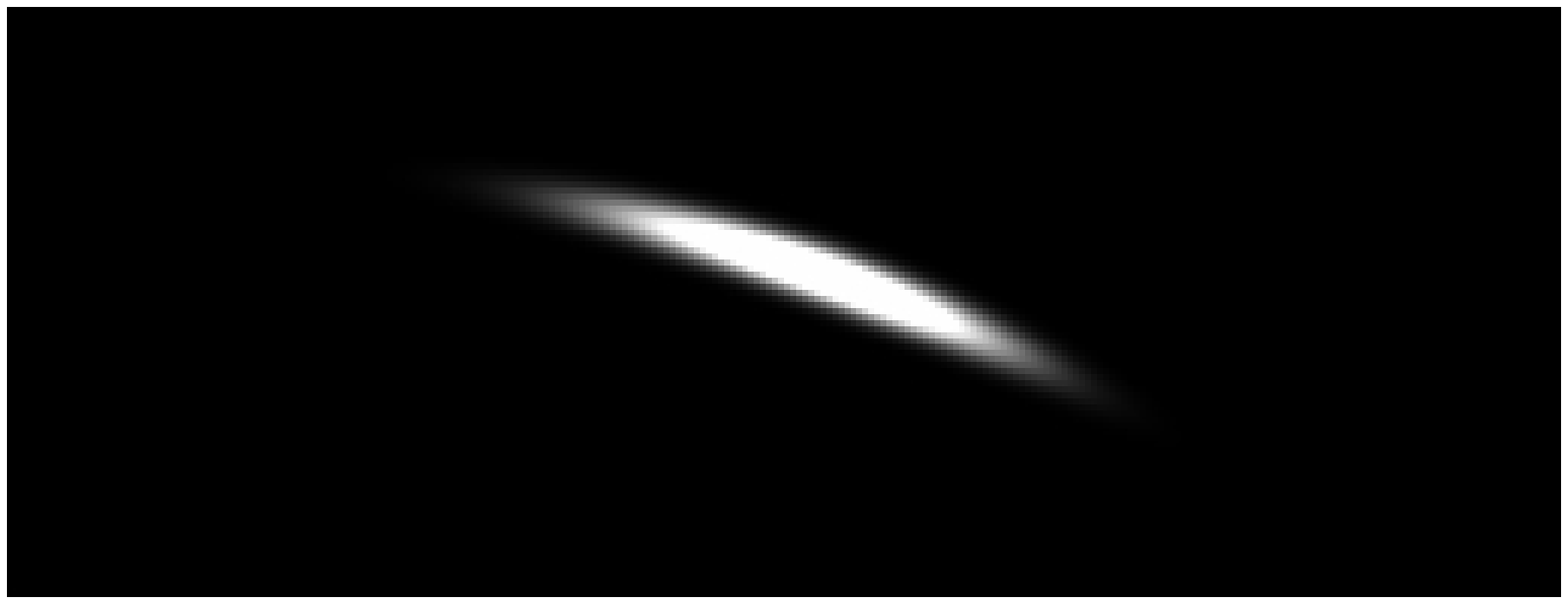}
\includegraphics[scale=0.25]{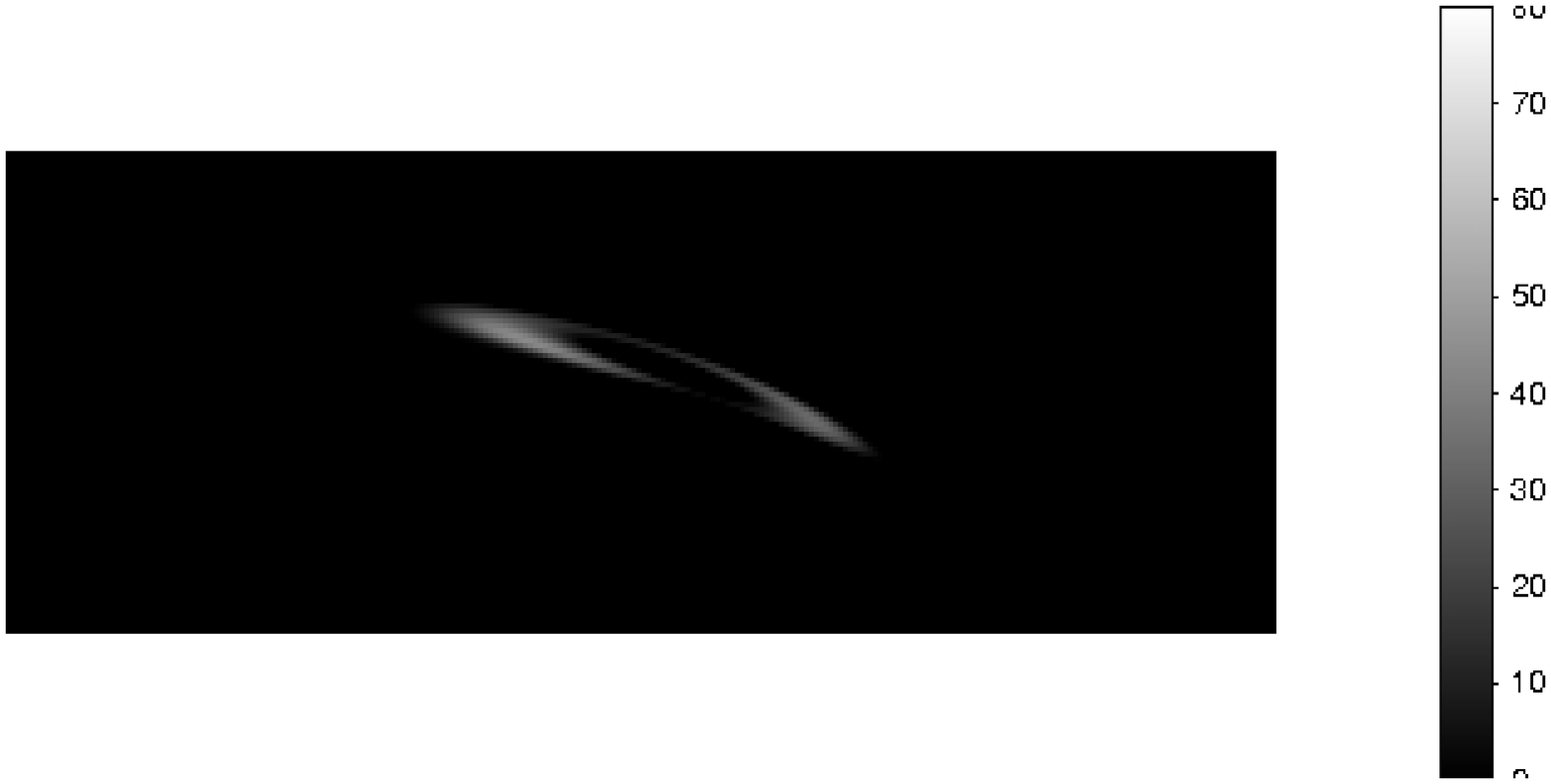}\\
\caption{ArcFitting results for AddArcs arcs. Left panels: original arc; middle panels: arc reproduced with PaintArcs using the parameters provided by the application of ArcFitting on original arcs; right panels: difference between original and reproduced arcs. }
\label{arcfitting_addarcs}
\end{minipage}
\end{center}
\end{figure*}

We ran ArcFitting on the arcs produced with AddArcs to infer their 
ArcEllipse+S\'ersic parameters and recreate them with PaintArcs. In 
Fig. \ref{arcfitting_addarcs} we show some examples of this application 
of the ArcFitting code. In the left panels we show the original simulated 
arcs (noiseless and without seeing convolution) from AddArcs. In the middle panel we show the ArcEllipse derived from 
ArcFitting and reproduced with PaintArcs. In the right panel we show the 
residual image resulting from the difference between the two previous ones. 
All panels use the same grayscale. Some substructures are visible in these
latter images, indicating the limitations of the ArcEllipse prescription. 
But the residuals are much less intense than the fitted arcs.

\begin{figure*}[!htb]
\begin{center}
\begin{minipage}[h]{1.0\linewidth}
\includegraphics[scale=0.4]{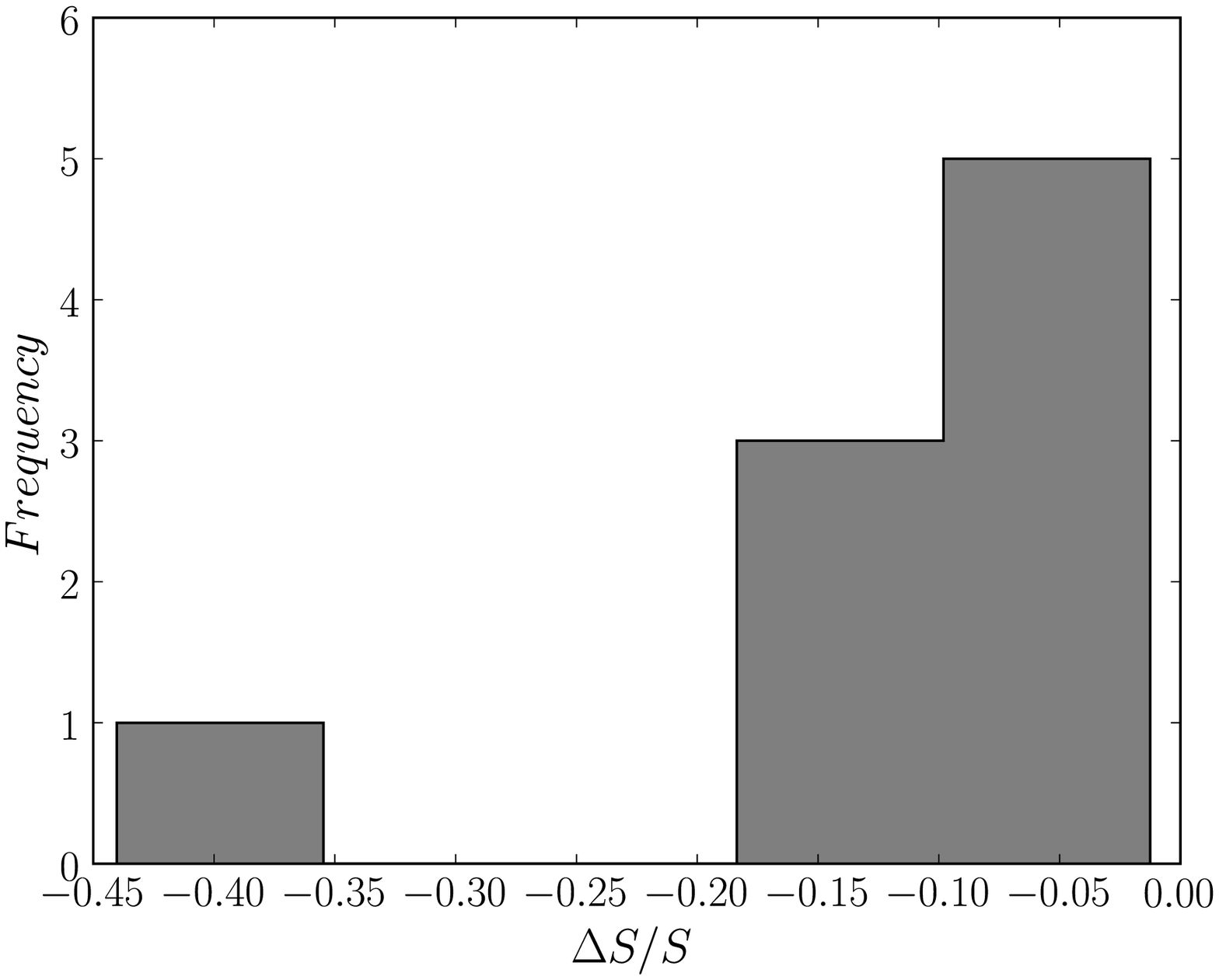}
\includegraphics[scale=0.4]{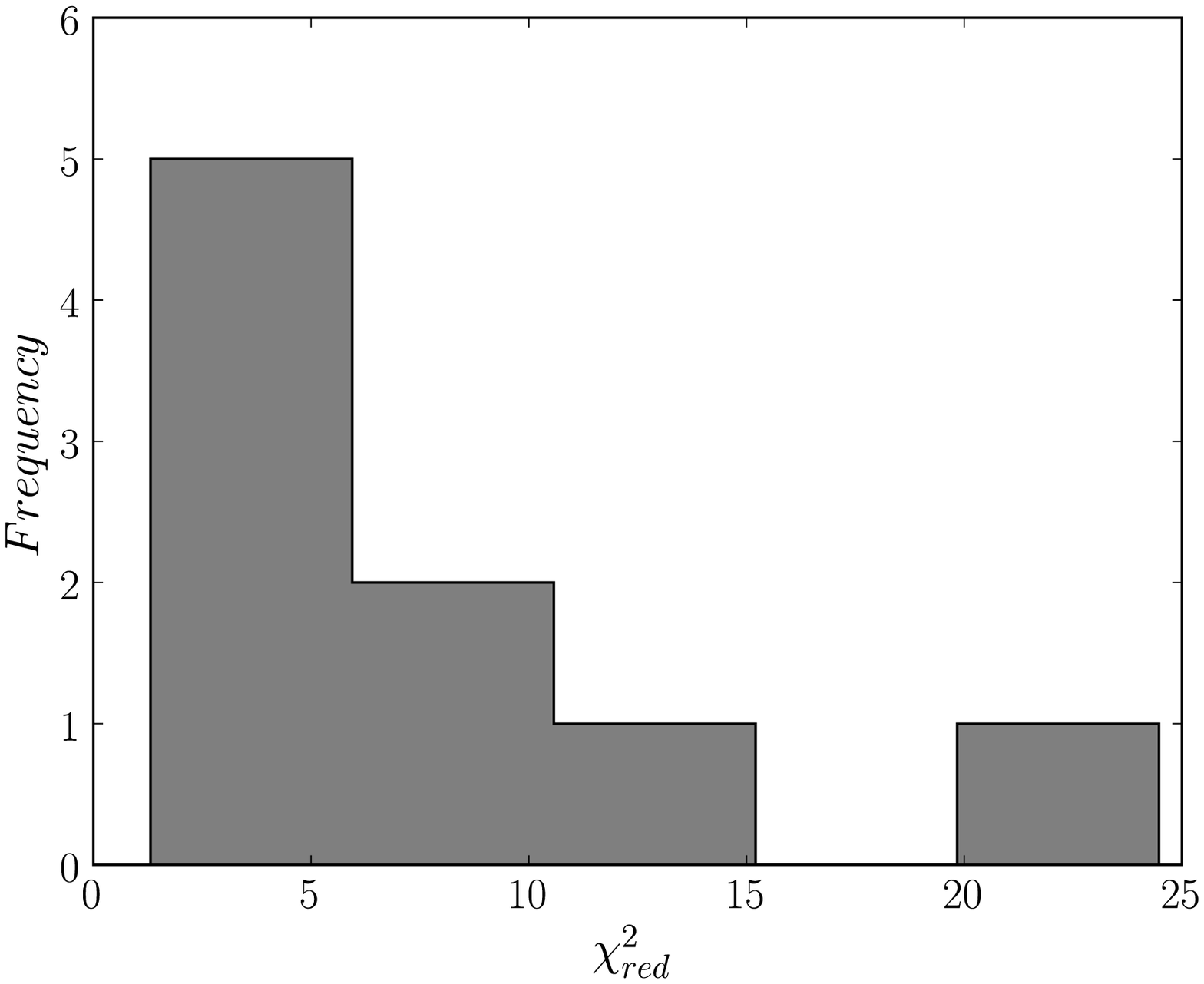}
\caption{Distribution of residual signal contrast $\Delta S / S$ (left panel) and of $\chi^2_{red}$ (right panel) resulting from applying ArcFitting on arcs simulated with AddArcs. }
\label{results_addarcs}
\end{minipage}
\end{center}
\end{figure*}

We computed the residual signal contrast $\Delta S / S$, as given in Eq. \ref{deltaS}, between each AddArcs simulation and its fitted ArcEllipse+S\'ersic. In the left panel of Fig. \ref{results_addarcs} we show the resulting distribution of $\Delta S / S$
values. The median $\Delta S / S$ value is $-0.09$, indicating that the 
ArcFitting process based on the ArcEllipse+S\'ersic model typically reproduces 
91\% of the signal of the input arc image. In the right panel of Fig. \ref{results_addarcs} we present the distribution of $\chi^2_{red}$ resulting from the application of ArcFitting on arcs simulated with AddArcs, whose median is $\chi^2_{red} = 5.27$. 

\subsection{Applying ArcFitting on arcs observed with the HST}
\label{hst_arcs}

We here test the reconstruction of HST arcs using the ArcFitting code based on 
the ArcEllipse+S\'ersic model. A sample of HST arcs was selected from the work 
by \citet{2005MNRAS.359..417S}. We chose bright and isolated arcs with 
spectroscopic redshift measurements that confirm their nature as lensed 
sources. A total of 12 arcs were selected. We use the same notation  here as 
the original authors. The arcs are C4, C8, and C12 from the Abell 68 galaxy 
cluster; H0, H1, H2, and H3 from Abell 963; B0/B1/B4 from Abell 383; M1 
and M3 from Abell 2218; and P0 and P2 from Abell 2219. Arc properties such 
as magnitude, $L/W$, and distance to the BCG of these arcs are presented 
in \citet{2005ApJ...627...32S}. 

\begin{figure}[!htb]
\begin{center}
\begin{minipage}[h]{1.0\linewidth}
\includegraphics[scale=0.23]{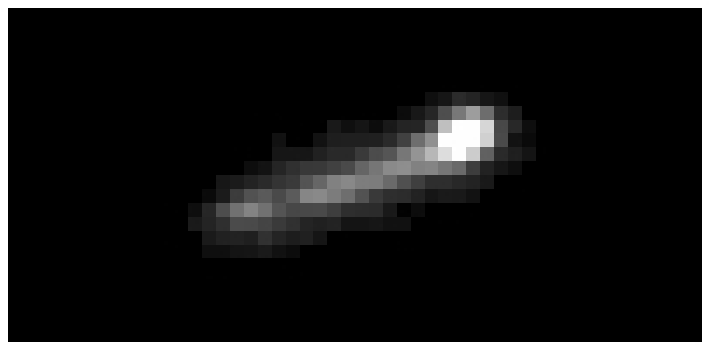}
\includegraphics[scale=0.23]{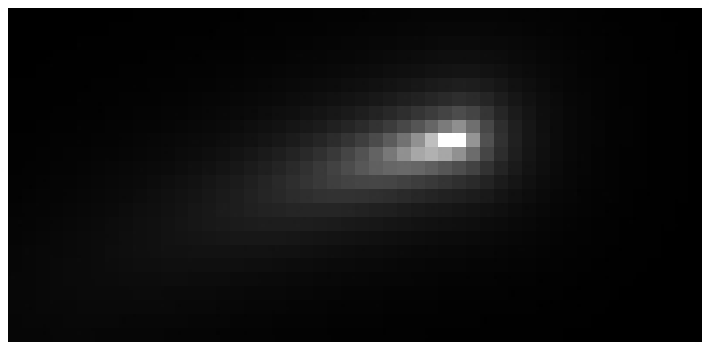}
\includegraphics[scale=0.23]{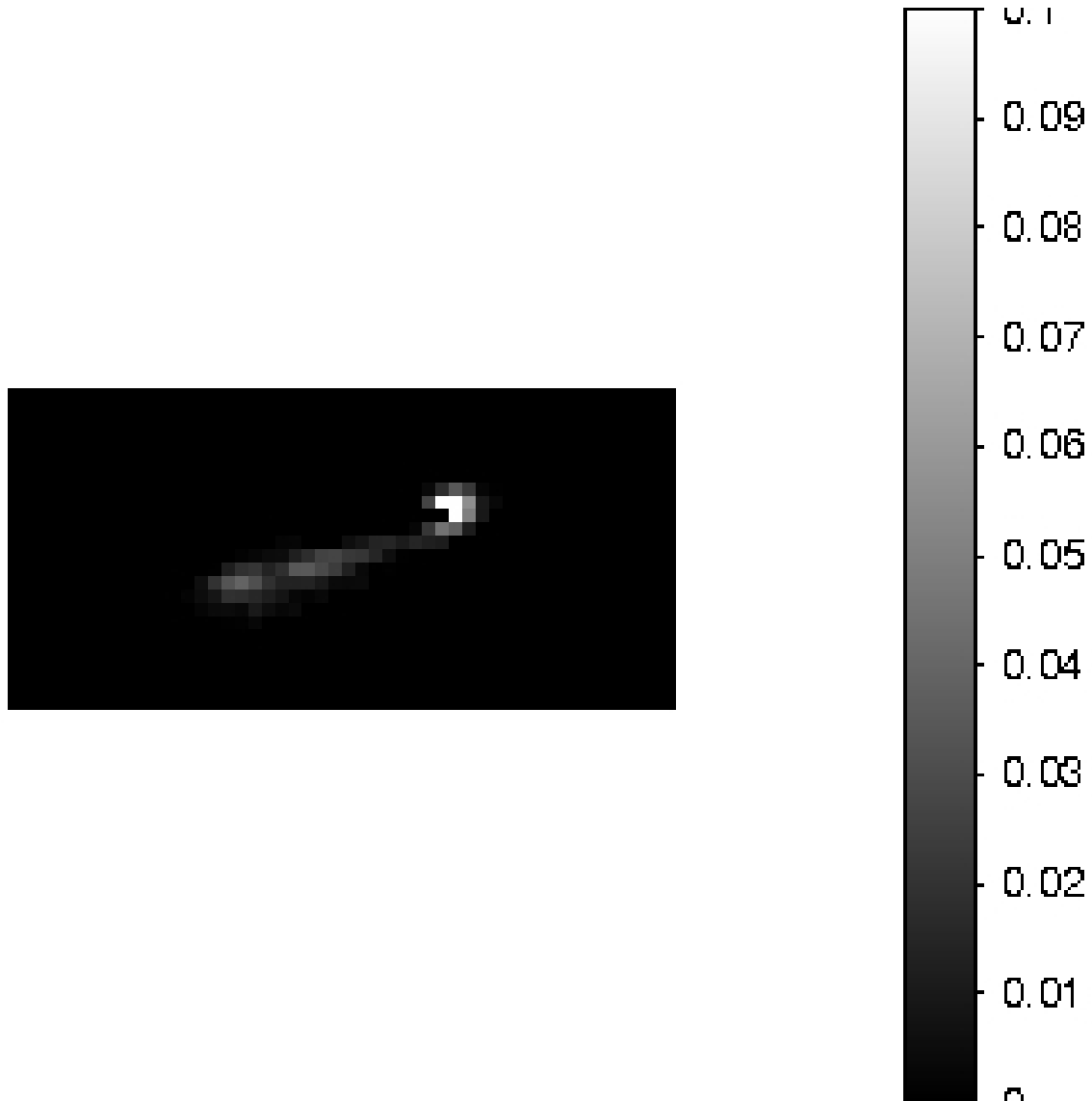}\\
\includegraphics[scale=0.23]{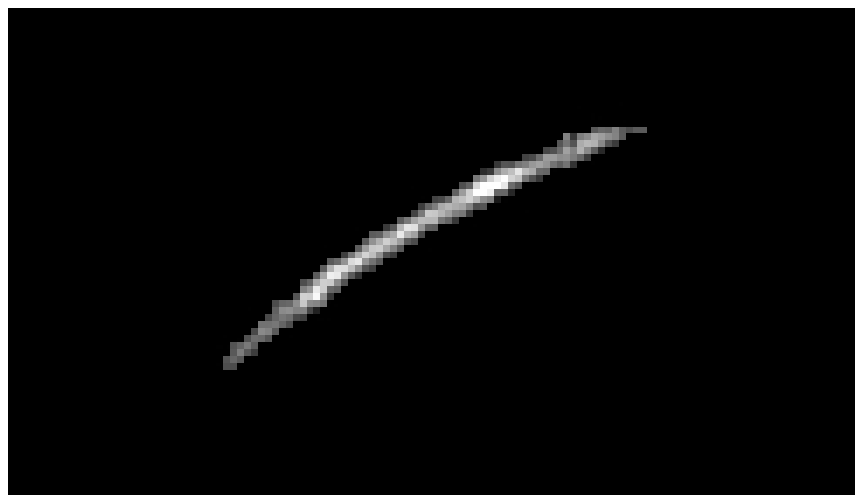} 
\includegraphics[scale=0.23]{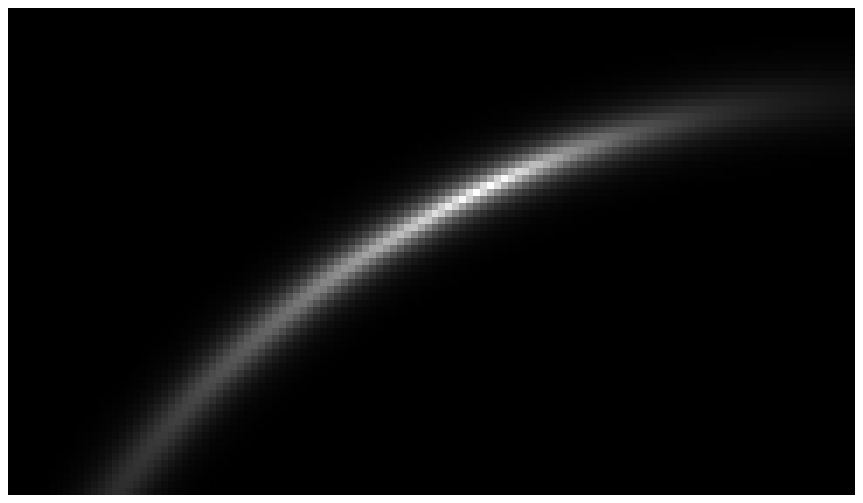}
\includegraphics[scale=0.23]{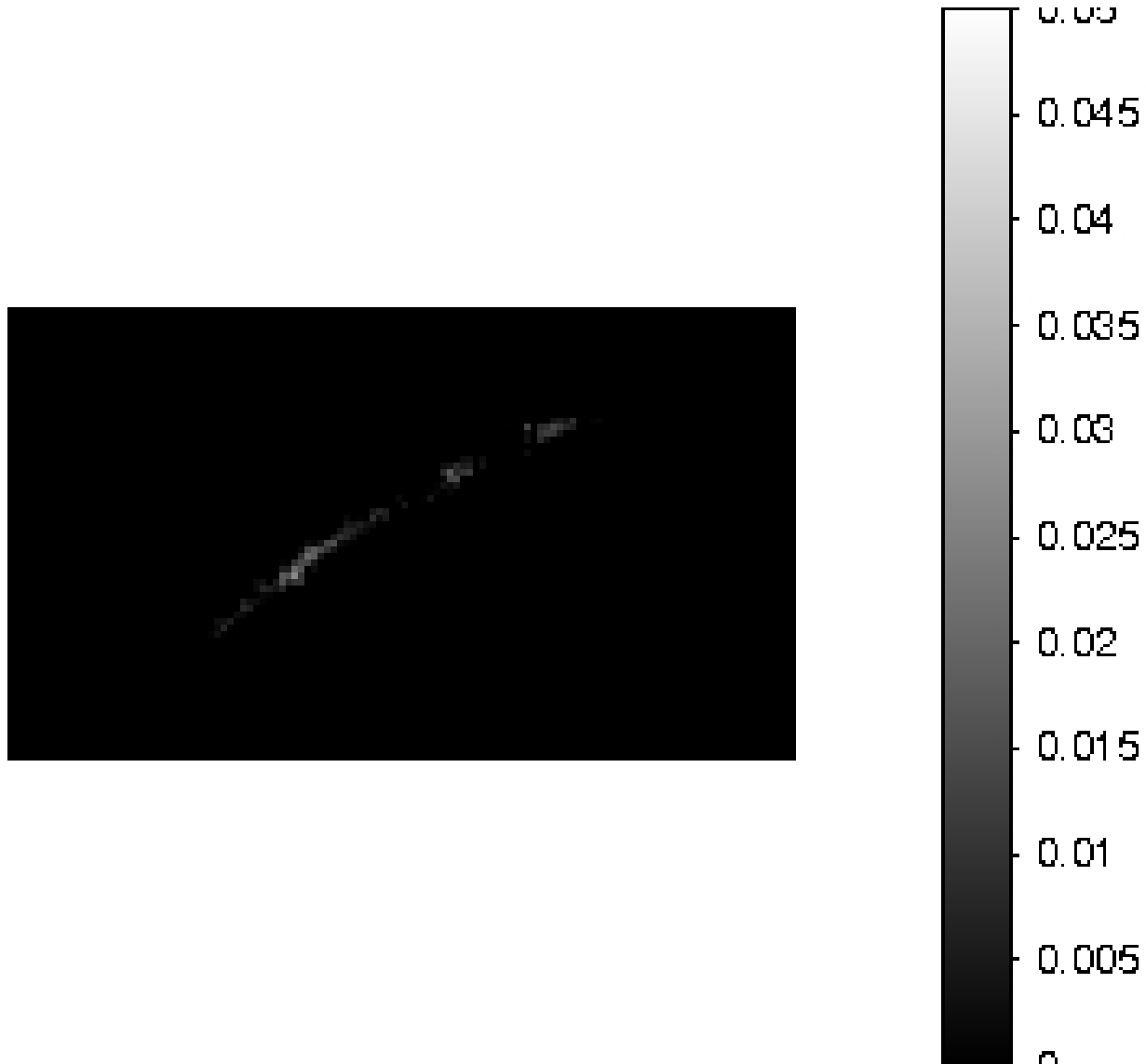}\\
\includegraphics[scale=0.23]{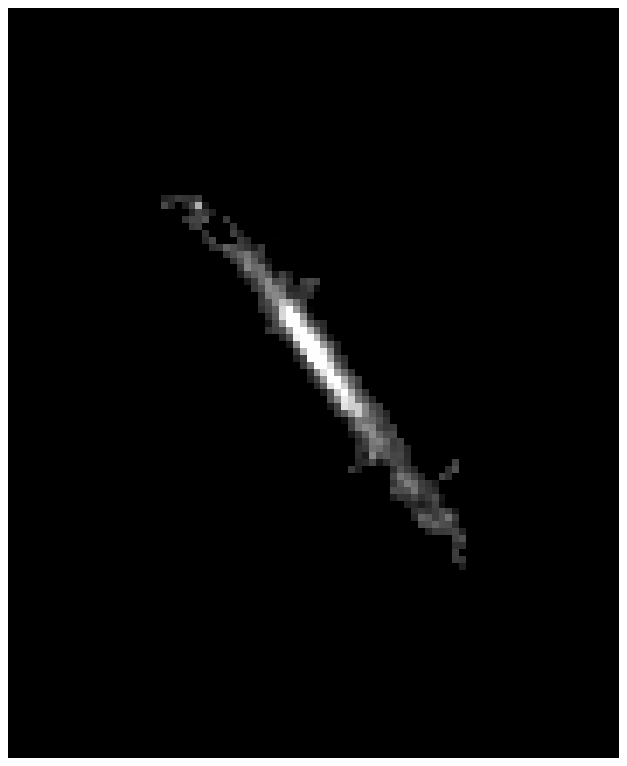} 
\includegraphics[scale=0.23]{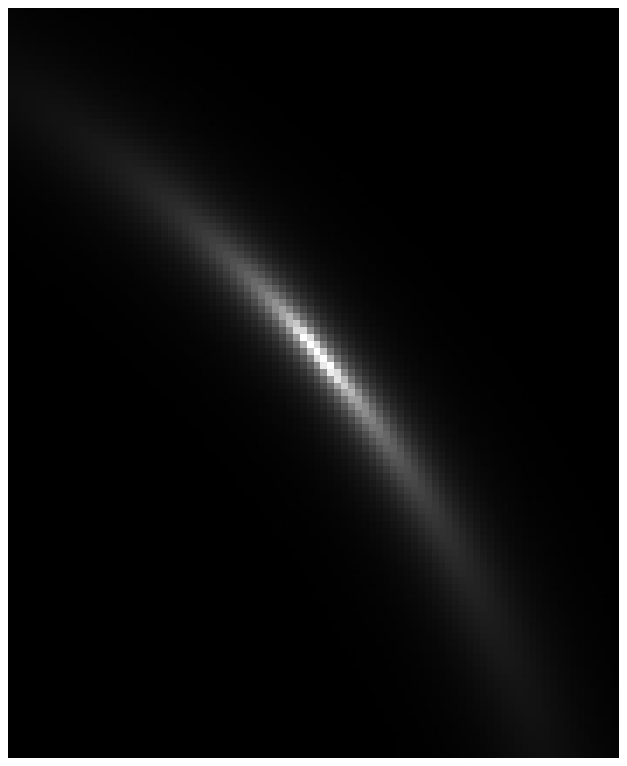}
\includegraphics[scale=0.23]{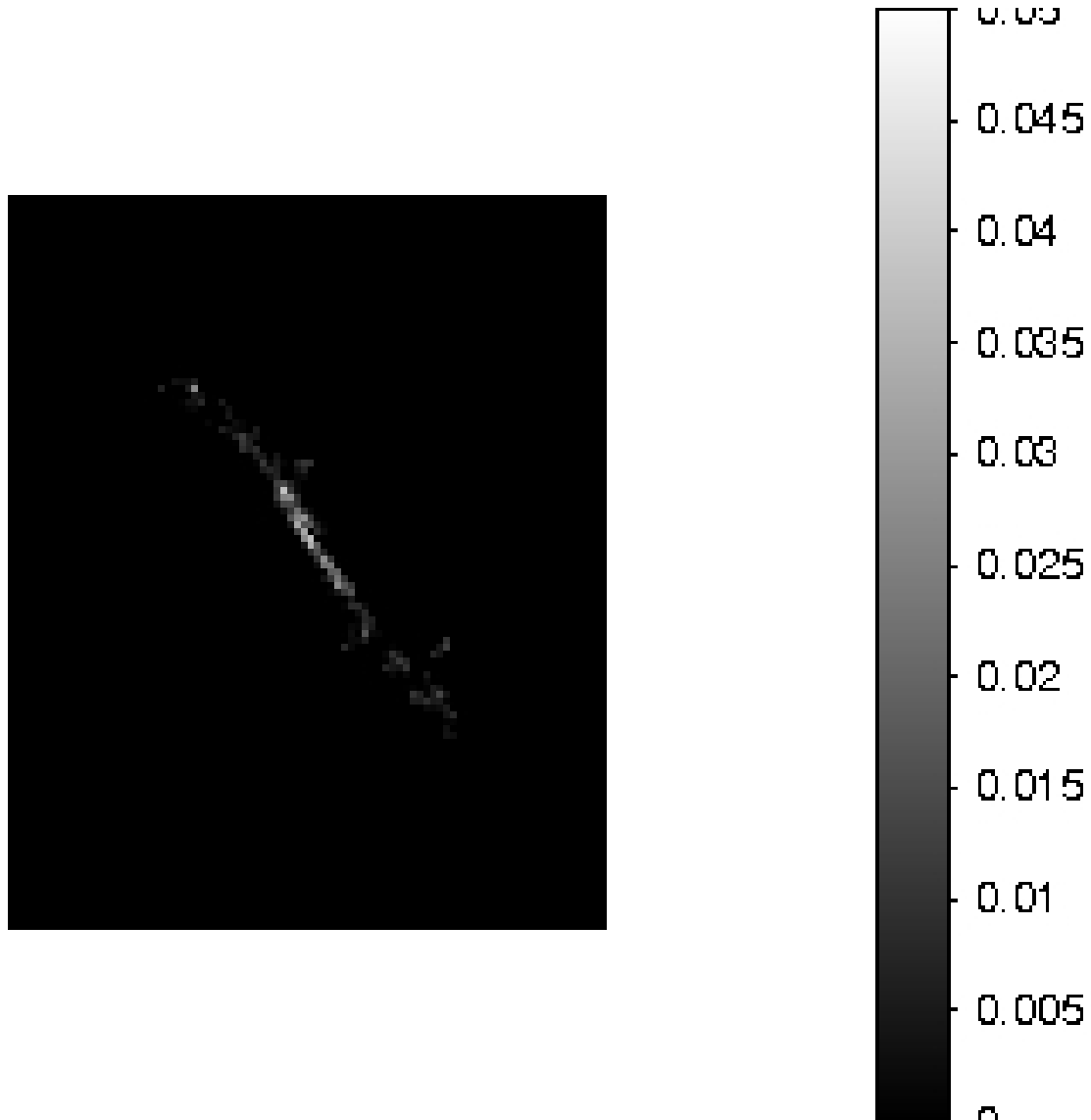}\\
\includegraphics[scale=0.23]{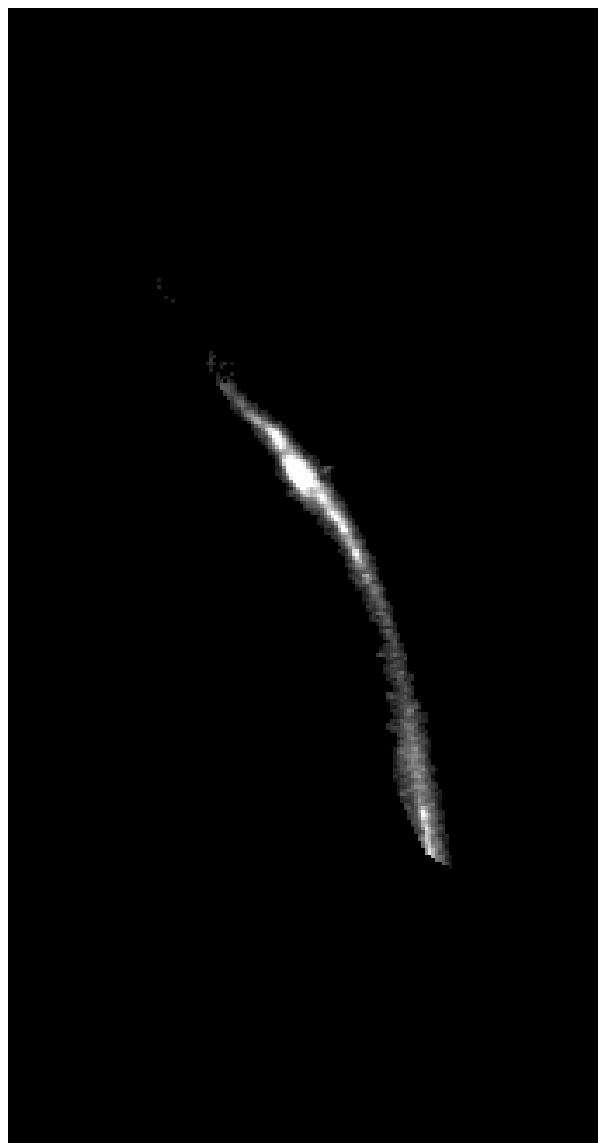} 
\includegraphics[scale=0.23]{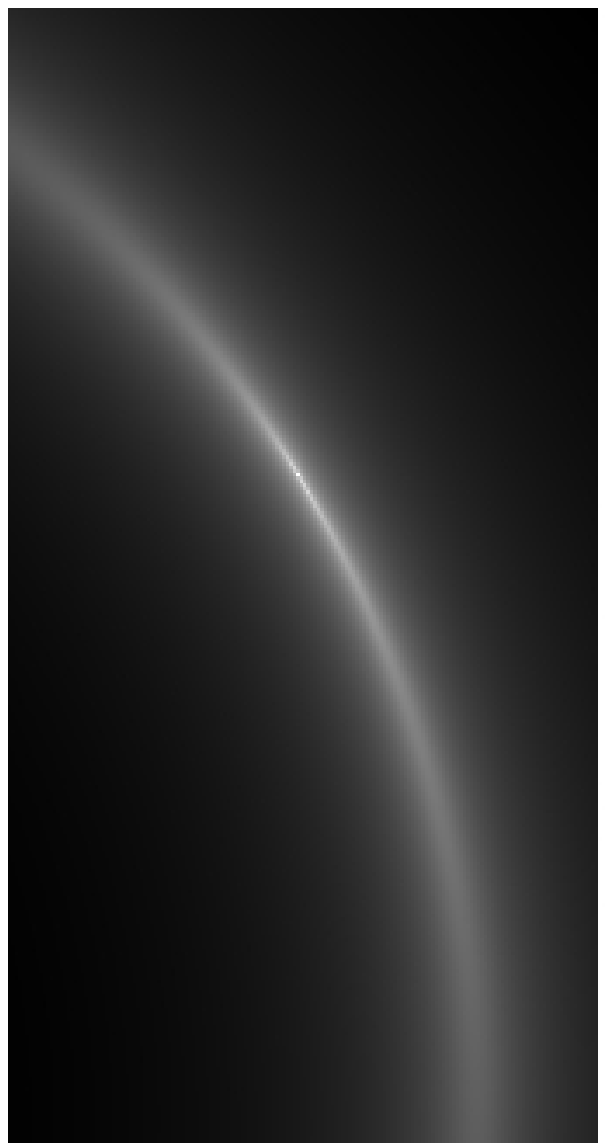}
\includegraphics[scale=0.23]{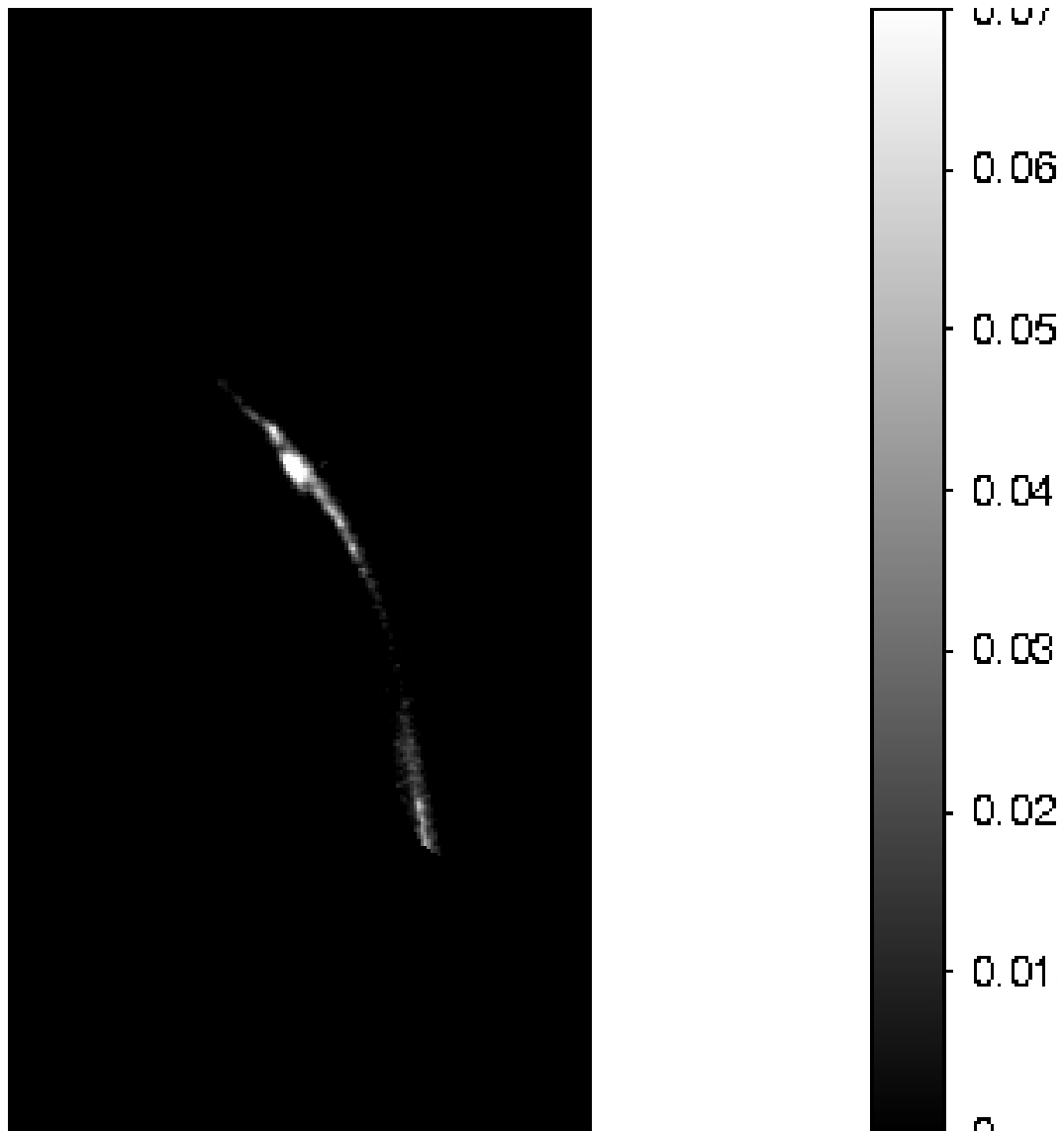}
\caption{Selected ArcFitting results for HST arcs. Left panels: original arc; middle panels: arc created with PaintArcs using the parameters derived from ArcFitting applied to the original arcs; right panels: difference between original and reproduced arcs. From top to bottom: arc C8 found in Abell 68; arc H0 of Abell 963; arc M3 of Abell 2218, and arc B0/B1/B4 of Abell 383.}
\label{hst_arcs}
\end{minipage}
\end{center}
\end{figure}

In Figure \ref{hst_arcs} we present some results of the application of 
ArcFitting to our selected HST arcs. The panels of Figure \ref{hst_arcs} are 
similar to those of Fig. \ref{arcfitting_addarcs}.
The left panels show the original HST arc images, the middle panels show the 
fitted ArcEllipse+S\'ersic model arcs, which was reproduced with PaintArcs, and 
the right panels show residual images. 

From the middle panels we can see that ArcFitting is efficient in recovering arc 
curvature and orientation. In some cases, as for the second and third arc 
in the figure, apart from the efficiency in recovering the arc shape, 
ArcFitting is also efficient in reproducing the arc brightness distributions, 
so that the residual images contain only little substructure and noise. 
However, for arcs with many substructures, such as the first and 
the last of that figure, ArcFitting cannot effectively reproduce the 
arc brightness profiles. 

\begin{figure*}[!htb]
\begin{center}
\begin{minipage}[b]{1.0\linewidth}
\includegraphics[scale=0.4]{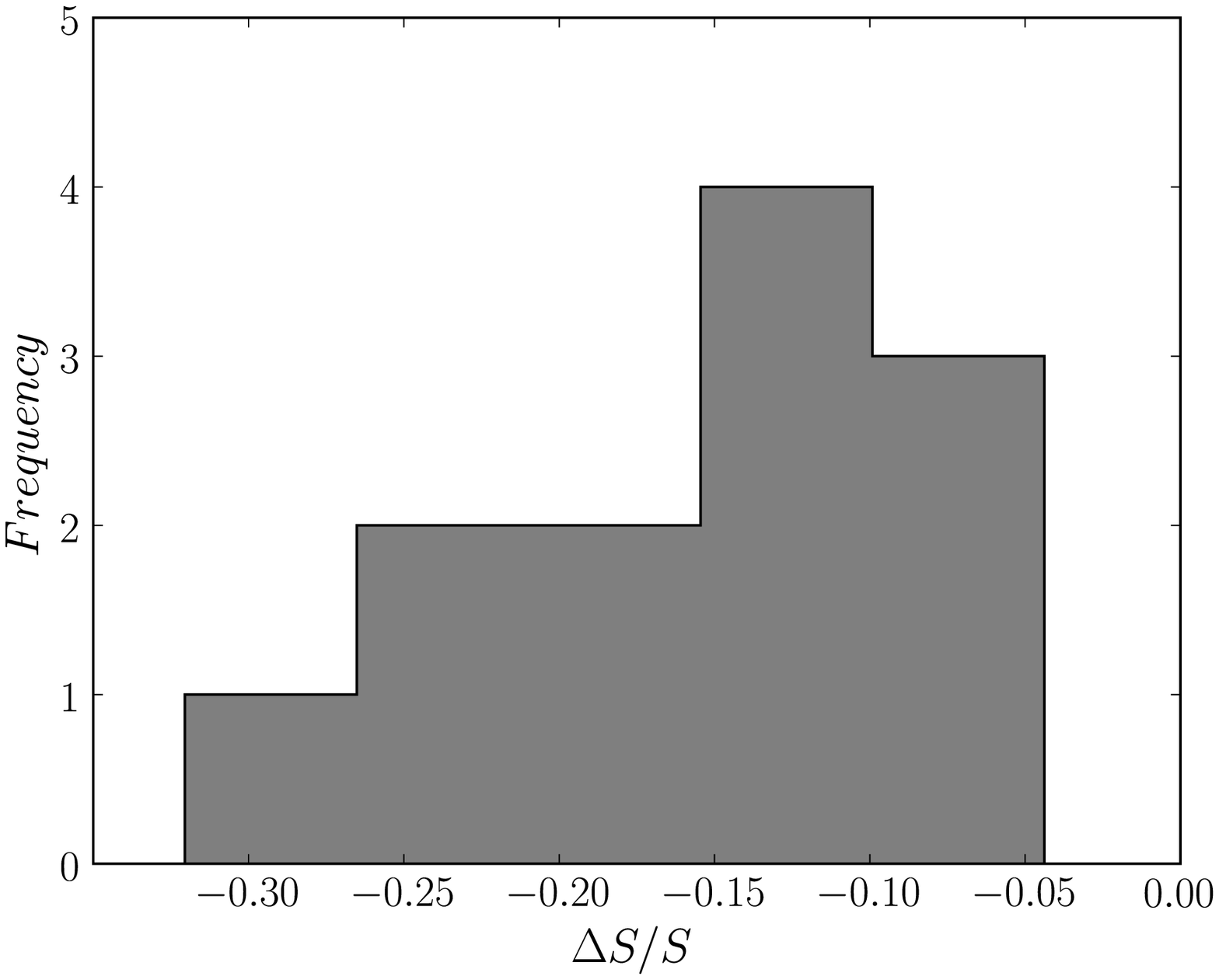}
\includegraphics[scale=0.4]{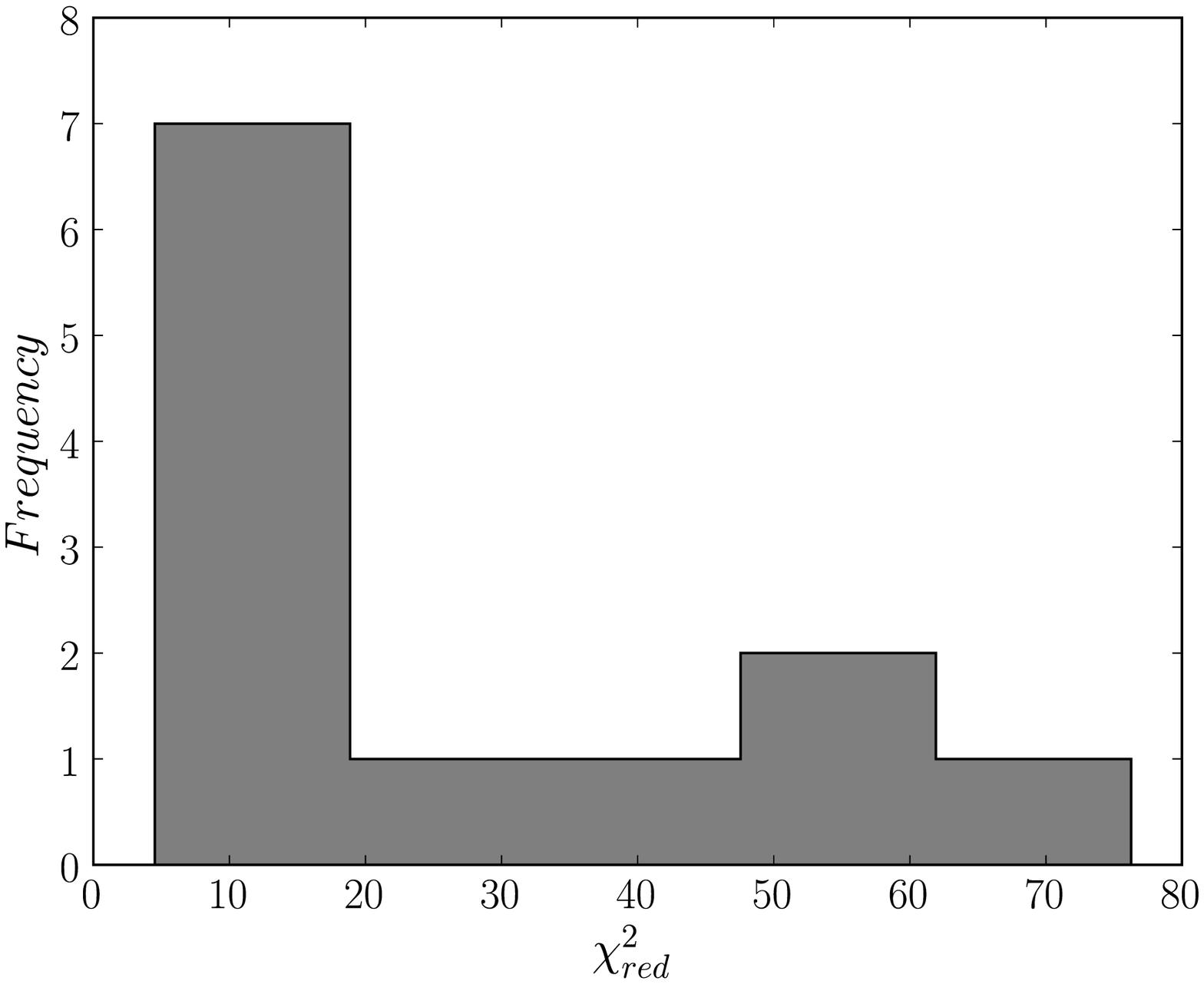}
\caption{Residual signal contrast ($\Delta S / S$) distribution (left panel) and $\chi^2_{red}$ distribution (right panel) resulting from applying ArcFitting on our selected sample of arcs from the HST.}
\label{results_hst}
\end{minipage}
\end{center}
\end{figure*}

We again used the residual signal contrast $\Delta S / S$, given by 
Eq. \ref{deltaS}, and the $\chi^2_{red}$ , given by Eq. \ref{chi2red2}, to assess the agreement between model and data. The 
distribution of $\Delta S / S$ values for all 12 arcs in our HST sample is 
shown in the left panel of Fig. \ref{results_hst}, while the distribution of $\chi^2_{red}$ is shown in the right panel of the same figure. The median of the $\Delta S / S$ distribution is $-0.13$, but values as high as $-0.32$ are found.
The median $\chi^2_{red}$ is $18.39$, but the distribution has a long tails 
toward much higher values. These results clearly attest that there is a stronger 
discrepancy between model and HST data than for the simulated AddArcs 
images. For all arcs, $\Delta S$ is negative, which means that ArcFitting is 
overestimating the arc signal. High values in both distributions correspond to the 
arcs that present many substructures (example: last arc of Fig. \ref{hst_arcs}, which has $\Delta S / S= -0.32$ and $\chi^2_{red} = 76.26$).

\begin{figure}[!htb]
\begin{center}
\includegraphics[scale=0.38]{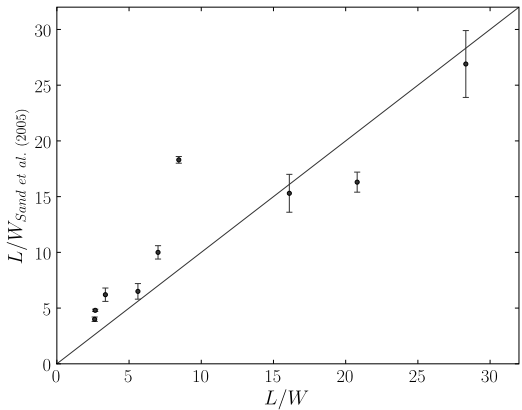}
\caption{Comparison between the 
$L/W$ measurements presented in \citet{2005ApJ...627...32S} and those obtained with 
ArcFitting. The solid line represents the identity function. }
\label{lw_hst}
\end{center}
\end{figure}

In Fig. \ref{lw_hst} we show a comparison of the
$L/W$ values measured by \citet{2005ApJ...627...32S} to those obtained with
ArcFitting. Since the arcs H1, H2, and H3 from Abell 963 are considered as a single object in \citet{2005ApJ...627...32S}, these arcs are not included in this comparison. No strong systematics is seen, which shows that ArcFitting provides a robust and automatic measure of $L/W$ ratio in real
gravitational arcs.

\subsection{Applying ArcFitting on arcs observed from the ground}
\label{cfhtls_arc}

We also ran the ArcFitting code to reconstruct arcs observed from the ground. We selected bright and isolated arcs in cluster and galaxy group lenses from the CFHTLS Strong Lensing Legacy Survey (SL2S), presented in the work by  \citet{2007A&A...461..813C}. Following the notation of the original authors, the seven selected arcs that compose our CFHTLS sample are SL2SJ021416-050315, SL2SJ021807-051536, SL2SJ085446-012137, SL2SJ141912+532612, SL2SJ142258+512440, SL2SJ143001+554647, and SL2SJ143140+553323. 

\begin{figure}[!htb]
\begin{center}
\begin{minipage}[b]{1.0\linewidth}
\includegraphics[scale=0.23]{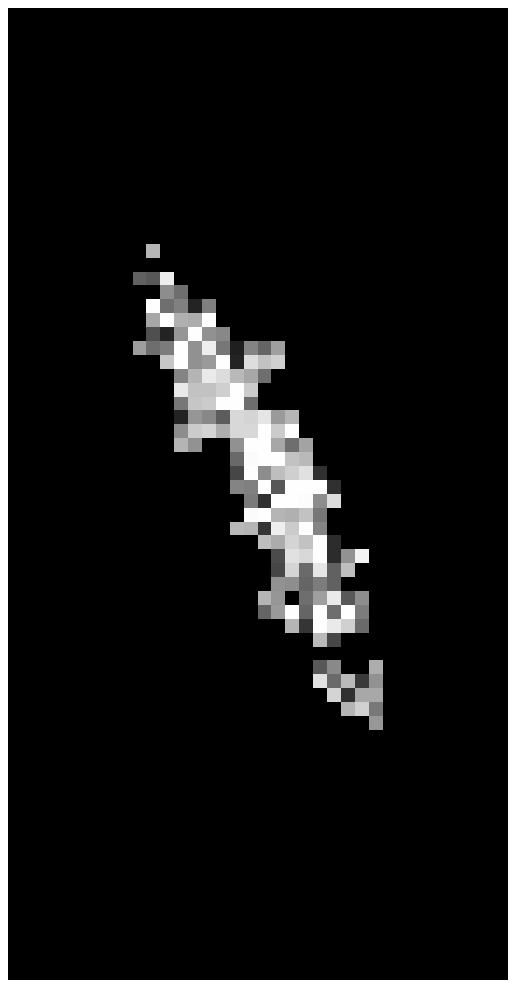}
\includegraphics[scale=0.23]{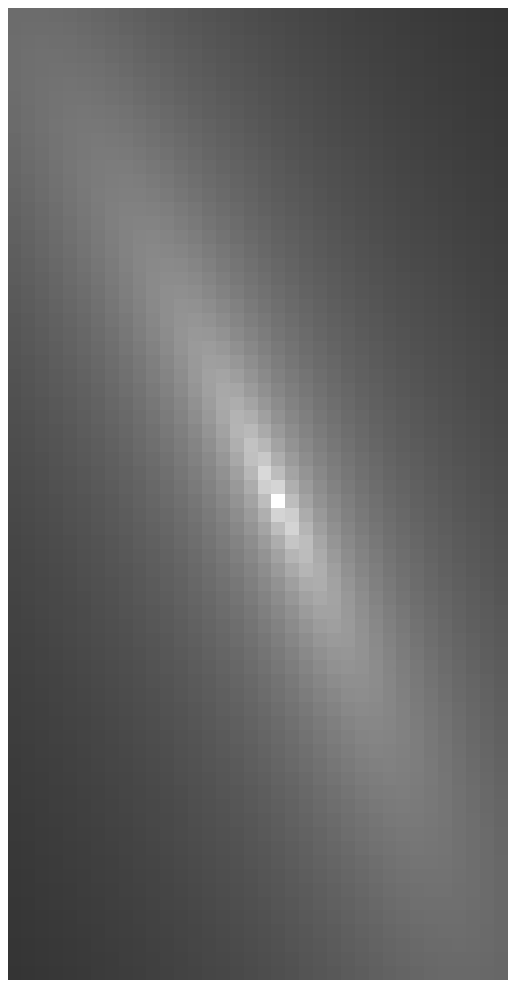}
\includegraphics[scale=0.23]{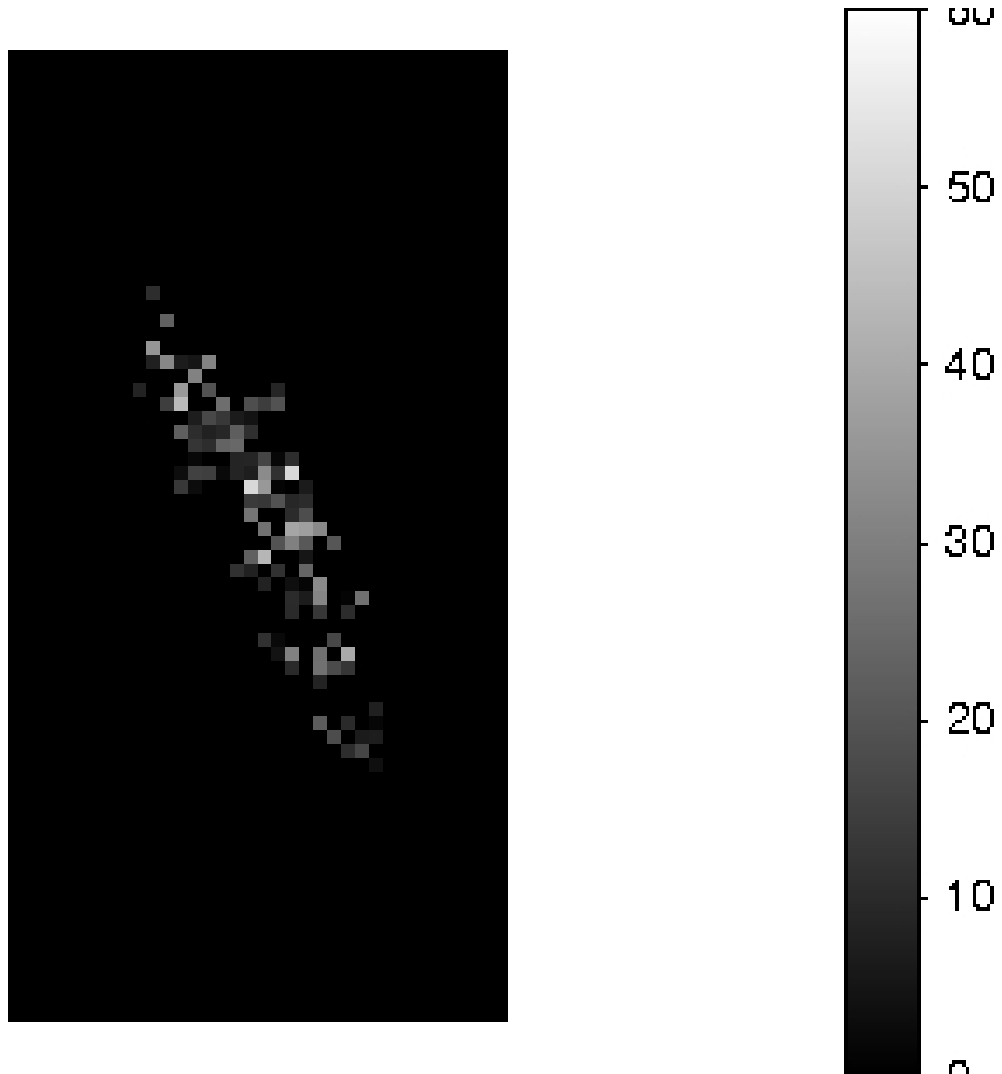}\\
\includegraphics[scale=0.23]{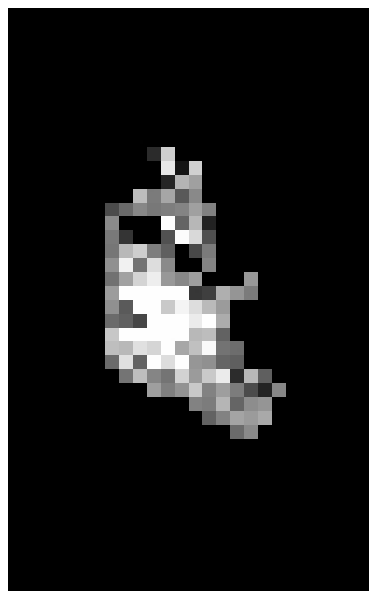} 
\includegraphics[scale=0.23]{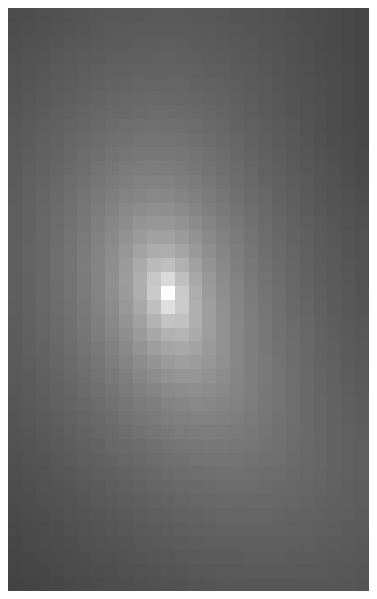}
\includegraphics[scale=0.23]{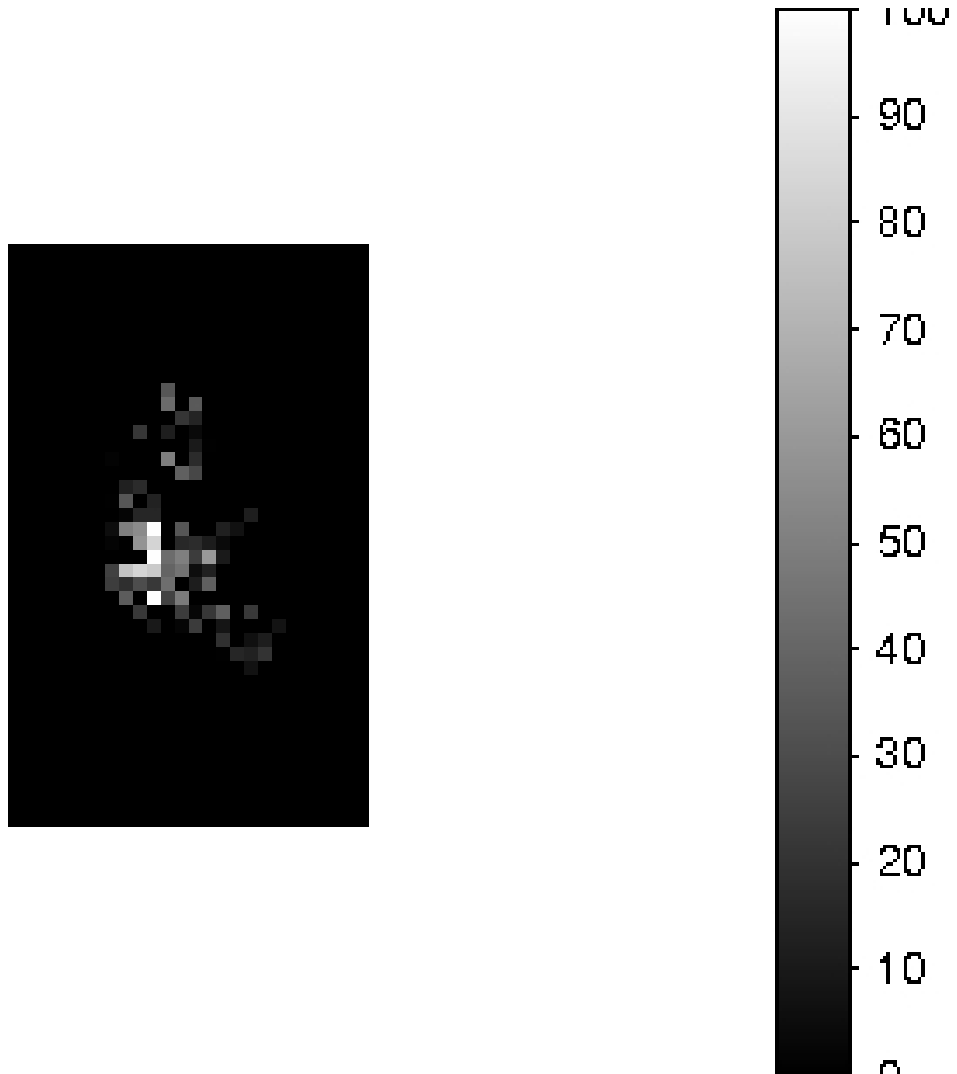}\\
\includegraphics[scale=0.23]{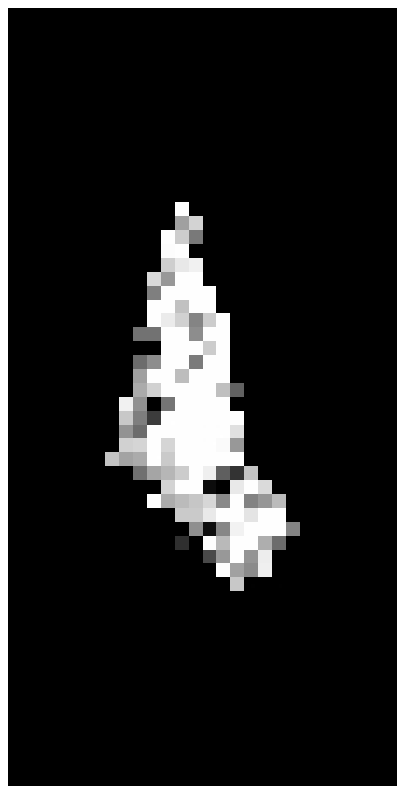}
\includegraphics[scale=0.23]{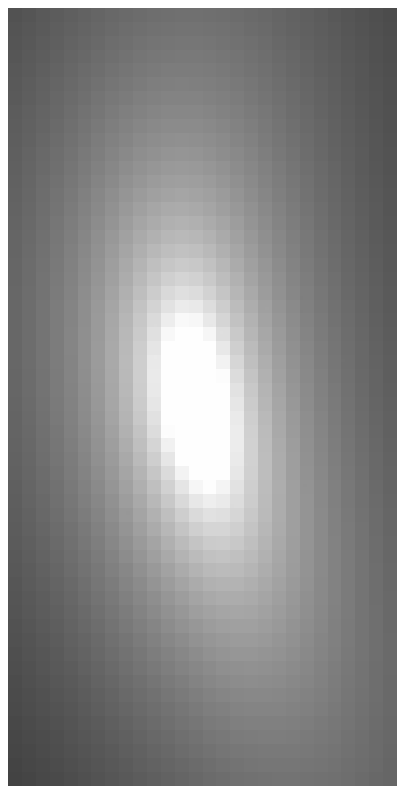}
\includegraphics[scale=0.23]{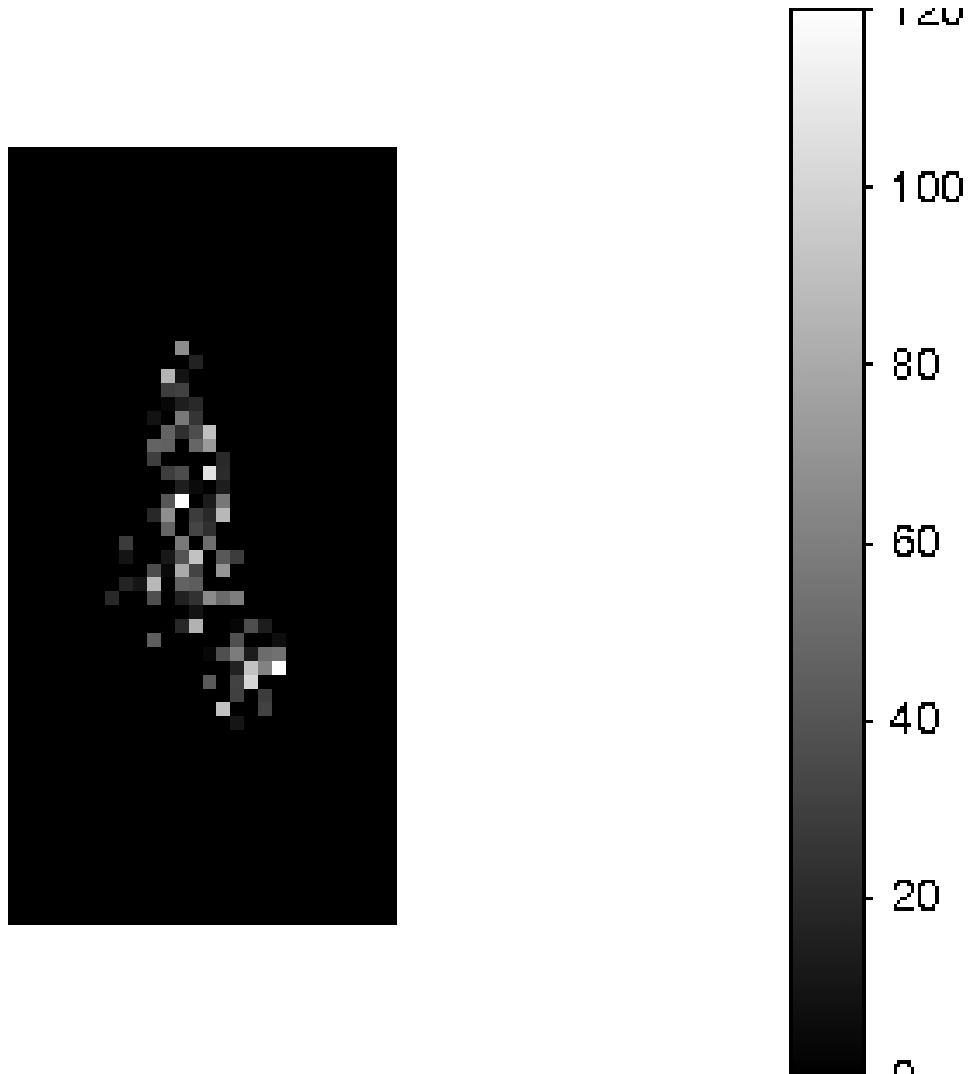}
\caption{ArcFitting results for CFHTLS arcs. Left panels: original arc; middle panels: ArcEllipse+S\'ersic image reproduced with PaintArcs using the parameters resulting from ArcFitting; right panels: difference between the previous two panels. From top to bottom: arcs SL2SJ141912+532612, SL2SJ143140+553323, and SL2SJ021807-051536.}
\label{cfhtls_arcs}
\end{minipage}
\end{center}
\end{figure}

The results of applying ArcFitting on three arcs of our CFHTLS sample are shown in Fig. \ref{cfhtls_arcs}. As in the previous comparisons, the left panels show the original arc images, the middle panels show the fitted ArcEllipse+S\'ersic model reproduced with PaintArcs, and the right panels show residual images 
resulting from the difference between the previous two. We can see that ArcFitting is efficient in recovering arc shape and brightness distributions.

\begin{figure*}[!htb]
\begin{center}
\begin{minipage}[b]{1.0\linewidth}
\includegraphics[scale=0.4]{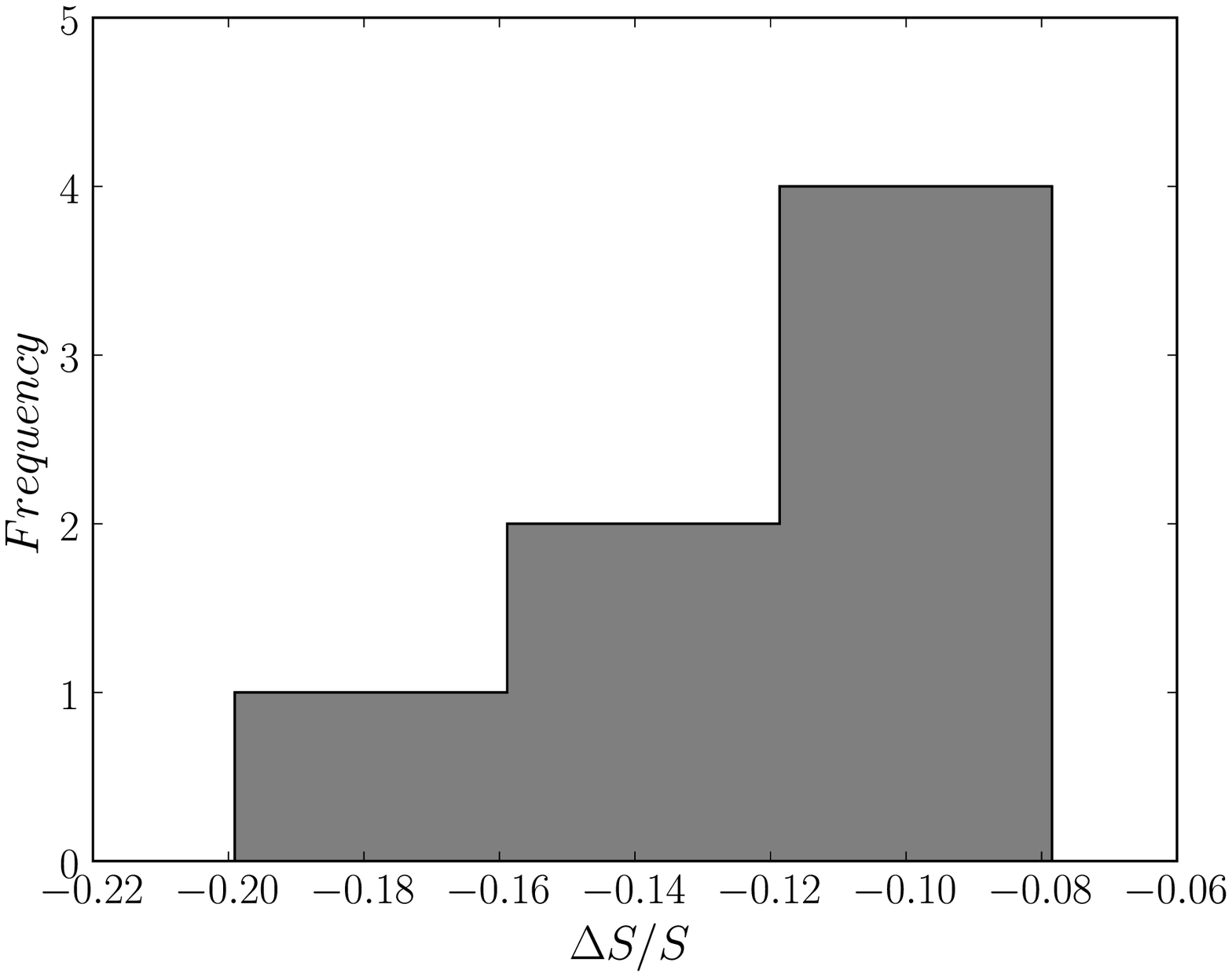}
\includegraphics[scale=0.4]{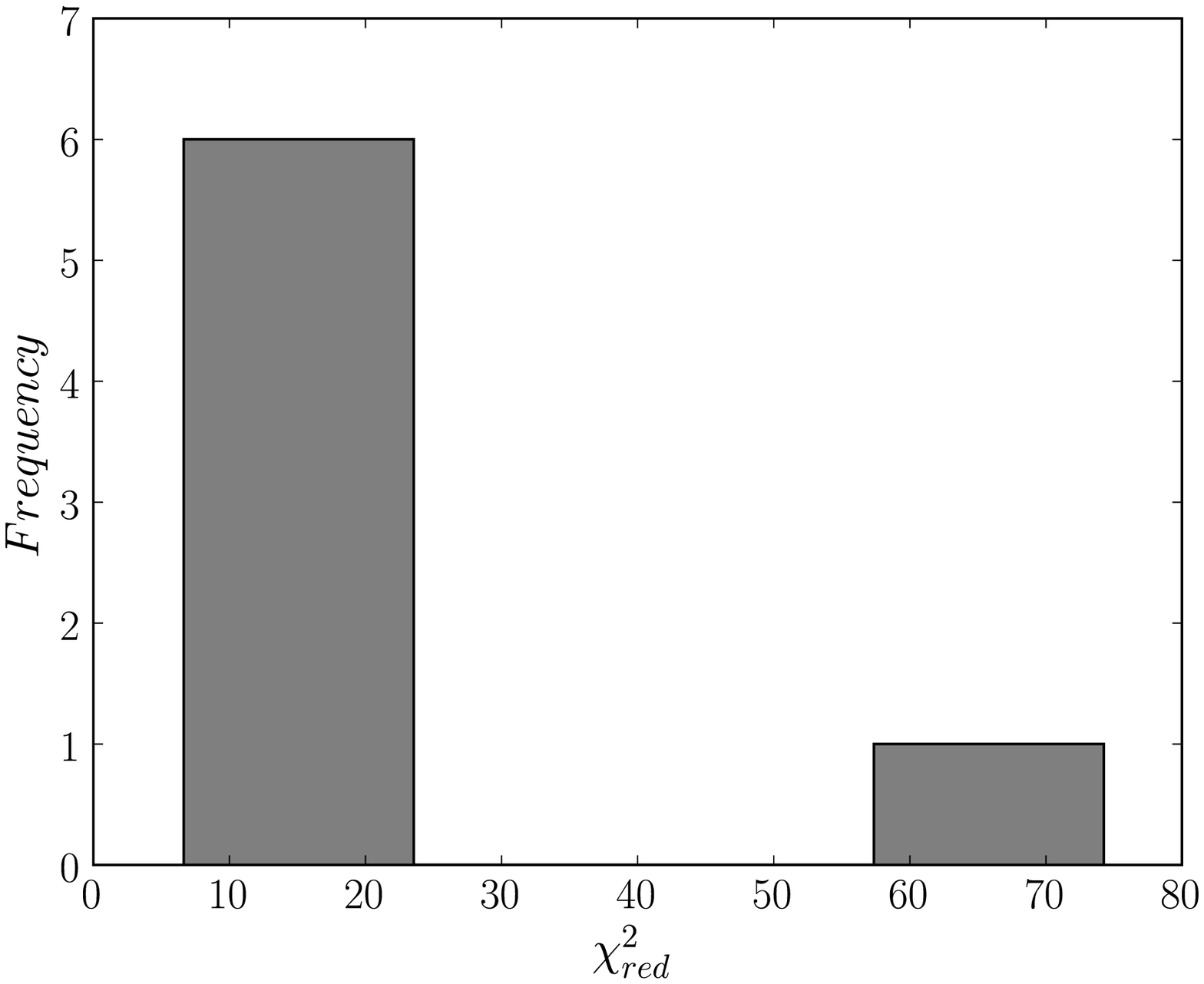}
\caption{Residual signal contrast ($\Delta S / S$) distribution (left panel) and $\chi^2_{red}$ distribution (right panel) resulting from applying ArcFitting on our selected sample of arcs from CFHTLS.}
\label{results_cfhtls}
\end{minipage}
\end{center}
\end{figure*}

In the left panel of Fig. \ref{results_cfhtls} we present the residual signal contrast distribution for the selected CFHTLS arc sample. As in the HST arcs, we can see that ArcFitting overestimates the arc signal, since $\Delta S < 0$ for all arcs of this sample. The median of the $\Delta S /S$ distribution is $-0.11$
and $\Delta S / S = -0.20$ in the worst case. In the rigth panel of the same figure, we show the distribution of $\chi^2_{red}$, whose median is $10.76$. In contrast to the HST arcs, no high $\chi^2_{red}$ tail is seen; the only outlier in the $\chi^2_{red}$ distribution corresponds to the arc SL2SJ085446-012137, the same arc that presents the highest value in the $\Delta S/S$ distribution.
The improvement in ArcFitting of CFHTLS arcs with respect to HST arcs is
likely caused by the smearing of the arc substructures caused
by seeing effects on the ground-based set.

\section{Conclusions and future perspectives}
\label{conclude}

We have presented and explored in detail the ArcEllipse, which is 
a simple analytical model that approximately reproduces the shape of real 
gravitational arcs. ArcEllipses are ordinary ellipses whose major axes are
bent along an arc of a circle and which therefore mimic the shape of tangential
arcs. They are fully characterized by a limited number of parameters
that quantify their curvature, length and width, degree of deformation,
and orientation. We also devised a simple prescription to incorporate
asymmetries to their shapes. This description of arc morphology is more comprehensive  than previously 
found in the literature. We also incorporated a S\'ersic light profile distribution to the
ArcEllipses in an attempt to explore a 
rough structural model for lensed galaxies.

We also implemented PaintArcs, which is a tool that 
currently generates simulated
ArcEllipse images, and ArcFitting, which is a code that derives model
parameters from images of real arcs. Both are Python codes built in a 
modular way, thus allowing flexibility. PaintArcs, for instance, 
incorporates the effects of pixelation, atmospheric seeing, and 
Poisson noise to the simulated ArcEllipse images, where the latter two can be turned on and off. 
ArcFitting is capable of deriving the model parameters
using different techniques that vary from simple and direct estimates 
applied on an image to statistical inference methods. Even though both 
codes currently adopt the ArcEllipse shapes and S\'ersic profiles 
for the arcs, other more complex
models can be added to their capabilities in the future.

We applied the ArcFitting code to different types of gravitational arcs. 
Realistic arc simulations based on ray-tracing techniques applied
to physical models of lens and source, such as those generated with Gravlens, have their
general features and total signal reasonably 
well fit by an ArcEllipse+S\'ersic model.
The resulting model from ArcFitting typically amounts to more than
90\% of the original signal from the arc. Typical reduced $\chi^2$ values
are on the order of $5$.

For real gravitational arcs around galaxy clusters, either imaged from
the ground or using the HST, the total signals are reproduced within 10-20\%
for ground-based images and within 10-30\% for HST images.
Comparing the $L/W$ values inferred from ArcFitting to those published for
HST data shows that ArcFitting successfully quantifies their axial ratio. We attribute the poorer fitting results 
from HST images to the presence of substructure in the arcs. This opens up the possibility of using ArcEllipse models to enhance these substructures, helping in identifying giant arcs formed from merging of multiple images. Extending this
reasoning, the ArcEllipse shape coupled with a single S\'ersic profile 
might in fact be used as a discriminator to filter out contaminanting 
elongated galaxies from a sample of arc candidates. Alternatively, 
gravitational arcs
formed from multiple merged images of the same source may be 
fit by more than one ArcEllipse+S\'ersic component to improve their
structural modeling.

Attempts to apply a full $\chi^2$ minimization fail in general, 
given the relatively
small number of degrees of freedom in the limited arc images available.
Our results also indicate that the ArcEllipse prescription is too simplistic
to yield significant improvement in the models based on 
formal statistical inference. We thus reached a compromise set-up in
ArcFitting by which statistical inference is applied only to the structural
S\'ersic parameters, while the shape ArcEllipse parameters are based on measurements over the arc images. The algorithm was succesfully validated in this way
by fitting a pure ArcEllipse+S\'ersic object with high precision.

The usefulness of the ArcEllipse+S\'ersic prescription should not be underestimated.
ArcEllipses of varying magnitudes, lenghts, $L/W$, and radii of curvature
can be speedily generated by PaintArcs and used in a number of arc studies,
taking advantage of the full control we have on the input arc parameters.
The Dark Energy Survey (DES) simulations have already included a sizable 
sample of arcs generated by PaintArcs by the DES strong lensing working group. 
Simulated objects using PaintArcs allow an efficient testing of 
{\it arcfinders}, which are codes meant to automatically detect gravitational 
arcs (Furlanetto et al., in prep.). 
In a 5,000 sq. deg. survey such as DES, usage of automated
{\it arcfinders} is unavoidable, since their large areas preclude visual
image inspection.

The ArcEllipse also allows an approximate initial characterization of observed
gravitational arcs. Even such rough morphological
and structural parameters of observed gravitational arcs are 
not commonly found in the literature. Our approach is thus providing both a qualitative and 
quantitative improvement on the amount of information extracted from observed gravitational arcs. The efficiency 
of ArcFitting is also an important aspect.
For large samples of arcs that contain perhaps thousands of them, as in
the case of the DES, an efficient first pass on arc parameters is a necessary tool
associated to the arcfinders themselves.

\section*{Acknowledgments}

We thank the anonymous referee for useful comments. We acknowledge the support of the Laborat\'orio Interinstitucional de e-Astronomia (LIneA) operated jointly by Centro Brasileiro de Pesquisas F\'isicas (CBPF), Laborat\'orio Nacional de Computa\c c\~ao Cient\'ifica (LNCC), and Observat\'orio Nacional (ON) and funded by the Minist\'{e}rio da Ci\^{e}ncia, Tecnologia e Inova\c{c}\~{a}o (MCTI) da Rep\'{u}blica Federativa do Brasil. The authors are supported by grants from the Conselho Nacional de Desenvolvimento Cient\'ifico e Tecnol\'ogico (CNPq) and Coordena\c{c}\~ao de Aperfei\c{c}oamento de Pessoal de N\'ivel Superior (CAPES). MM is also partially supported by FAPERJ (grant E-26/110.516/2012). MM acknowledges the hospitality of Instituto Balseiro at Centro Atomico Bariloche where part of this work was done.

This paper makes use of observations obtained at the Southern Astrophysical Research (SOAR) telescope, which is a joint project of MCTI, the U.S. National Optical Astronomy Observatory (NOAO), the University of North Carolina at Chapel Hill (UNC), and Michigan State University (MSU); 
observations made with the NASA/ESA Hubble Space Telescope, obtained from the ESO/ST-ECF Science Archive Facility; and observations obtained with MegaPrime/MegaCam, a joint project of CFHT and CEA/DAPNIA, at the Canada-France-Hawaii Telescope (CFHT) which is operated by the National Research Council (NRC) of Canada, the Institut National des Science de l'Univers of the Centre National de la Recherche Scientifique (CNRS) of France, and the University of Hawaii. This work is based in part on data products produced at TERAPIX and the Canadian Astronomy Data Centre as part of the Canada-France-Hawaii Telescope Legacy Survey, a collaborative project of NRC and CNRS.

\bibliographystyle{aa} 
\bibliography{refs} 

\end{document}